\documentclass{article}

\usepackage{arxiv}

\usepackage[utf8]{inputenc} 
\usepackage[T1]{fontenc}    
\usepackage{hyperref}       
\usepackage{url}            
\usepackage{booktabs}       
\usepackage{amsfonts}       
\usepackage{nicefrac}       
\usepackage{microtype}      
\usepackage{lipsum}		
\usepackage{graphicx}
\usepackage{doi}
\usepackage{bm}
\usepackage{amssymb}
\usepackage{float}
\usepackage{amsmath}
\DeclareMathOperator{\asinh}{asinh}
\DeclareMathOperator{\atanh}{atanh}
\usepackage{physics}

\usepackage[table,xcdraw]{xcolor}

\title{Dynamic dielectric function and phonon self-energy from electrons strongly correlated with acoustic phonons in 2D Dirac crystals}

\date{November 1, 2021}	

\author{ \href{}{\hspace{1mm}Sina Kazemian}\\
	Department of Physics and Astronomy\\
	The University of Western Ontario\\
	London, ON, Canada \\
	\texttt{skazemi5@uwo.ca } \\
	\And
\href{}{Giovanni Fanchini} \\
	Department of Physics and Astronomy\\
	The University of Western Ontario\\
	London, ON, Canada \\
	\texttt{gfanchin@uwo.ca} \\
}

\renewcommand{\shorttitle}{Dielectric response function and phonon self-energy in 2D Dirac crystals}

\hypersetup{
pdftitle={A template for the arxiv style},
pdfsubject={q-bio.NC, q-bio.QM},
pdfauthor={Ghazal Farhani, Alexander Kazachek, Boyu Wang},
pdfkeywords={First keyword, Second keyword, More},
}

\begin{document}
\maketitle

\begin{abstract}
The unique structure of two-dimensional (2D) Dirac crystals, with electronic bands linear in the proximity of the Brillouin-zone boundary and the Fermi energy, creates anomalous situations where small Fermi-energy perturbations critically affect the electron-related lattice properties of the system. The Fermi-surface nesting (FSN) conditions determining such effects via electron-phonon interaction, require accurate estimates of the crystal’s response function $(\chi)$ as a function of the phonon wavevector \textbf{q} for any values of temperature, as well as realistic hypotheses on the nature of the phonons involved. Numerous analytical estimates of $\chi(\textbf{q})$ for 2D Dirac crystals beyond the Thomas-Fermi approximation have been so far carried out only in terms of dielectric response function $\chi(q,\omega)$, for photon and optical-phonon perturbations, due to relative ease of incorporating a q-independent oscillation frequency $(\omega)$ in their calculation. However, models accounting for Dirac-electron interaction with ever-existing acoustic phonons, for which $\omega$ does depend on \textbf{q} and is therefore dispersive, are essential to understand many critical crystal properties, including electrical and thermal transport. The lack of such models has often led to the assumption that the dielectric response function $\chi(q)$ in these systems can be understood from free-electron behavior. Here, we show that different from free-electron systems, $\chi(q)$ calculated for acoustic phonons in 2D Dirac crystals using the Lindhard model, exhibits a cuspidal point at the FSN condition even in the static case and at 0 K. Strong variability of $\frac{\partial\chi}{\partial q}$ persists also at finite temperatures, while $\chi(q)$ may tend to infinity in the dynamic case even where the speed of sound is small, albeit nonnegligible, over the Dirac-electron Fermi velocity. The implications of our findings for electron-acoustic phonon interaction and transport properties such as the phonon line width derived from the phonon self-energy will also be discussed.

\end{abstract}

\maketitle


\section{\label{sec:level1}Introduction}

Dirac crystals are a broad class of zero band-gap solids in which the electronic band structure is linear in the crystal momentum, $\hbar\textbf{k}$, instead of quadratic, as commonly observed in metals and semiconductors.\cite{wehling2014dirac,wang2015rare,cayssol2013introduction} This leads to dispersionless electrons with a behavior reminiscent of photons on the Dirac light cone in relativity.\cite{dirac1928quantum} In two-dimensional (2D) Dirac crystals, the best known of which is graphene, the density of electronic states near the Dirac point also linearly tends to zero, thus enabling additional remarkable properties, including extreme carrier mobility and quantum Hall effects, and the possibility of topologically  insulating characteristics.\cite{xu2016hydrogenated,offidani2018anomalous,zhang2021two,kou2013graphene} The experimental discovery of a plethora of new 2D Dirac crystals in recent years \cite{chowdhury2016theoretical,acun2015germanene,meng2021two,wang2019review,cayssol2013introduction, moore2010birth} compels their deeper understanding.

In undoped Dirac crystals, the valence and conduction bands meet at the Brillouin-zone boundary Dirac \textbf{K}-point. The Fermi energy $E_{\mathrm{F}}$ coincides with the energy of this point, with the Fermi surface area collapsing to zero. Thus, the crystal’s Fermi surface degenerates into a single point of the electronic band structure.\cite{wehling2014dirac,wang2015rare} Such a degeneracy enables us to tune the density of electronic states at $E_{\mathrm{F}}$, as well as the system’s electrical and thermal conductivity, via external electric fields or tunable doping–an effect that leads the carrier density to increase by orders of magnitude upon small fluctuations of $E_{\mathrm{F}}$.\cite{novoselov2007electronic,li2018review} These Fermi-level shifts are also expected to dramatically alter the strength of the interaction between charge carriers and lattice phonons,\cite{roy2014migdal,hu2021phonon} which may have profound effects on the applicability of the Born-Oppenheimer approximation, leading to highly specific dielectric properties from electrons strongly correlated with acoustic phonons in 2D Dirac crystals.\cite{pisana2007breakdown,born1927quantentheorie,kazemian2017modelling} 

A key parameter to understand the effects of strong electron-phonon correlation is the crystal’s dielectric response function, $\chi(q)$. $\chi(q)$ differs from the dielectric susceptibility $\chi(q,\omega)$ in that it takes into account the dispersion relationship $\omega_{\textbf{q}}$ of the phonons for which it is calculated and cannot simply be obtained from replacing $\omega_{\textbf{q}}$ into $\chi(q,\omega)$, with which processes violating the conservation of energy or momentum would be inappropriately considered. For phonons of wavevector, $\textbf{q}$, $\chi(q)$ originates from the superposition of all of the inelastic scattering of electrons and holes by such phonons. The customary approach for calculating $\chi(q)$ with appropriate selection rules relies on the Thomas-Fermi (also known as) Debye-Huckel \cite{jishi2013feynman,ashcroft2001festkorperphysik,huckel1923theory} approximation, which assumes long wavelength and a short wave number from the scattered phonons. Such an assumption is specifically designed for metals with large $(>10^{22} cm^{-3})$ electron densities and large Fermi surfaces in the proximity of $E_{\mathrm{F}}$ and, therefore, q much shorter than the dimensions of the Fermi surface, implying short screening length.\cite{maldague1978many} On the contrary the Fermi surface area in undoped 2D Dirac crystals collapses to zero, as already stated. The screening length may tend to infinity and the Fermi wavenumber $\textbf{k}_\mathrm{F}$ corresponding to the Fermi surface radius, is expected to remain very small also in the presence of external electric fields and doping, always leading to free-electron densities $10^{19} cm^{-3}$ orders of magnitude below the expected applicability of the models \cite{huckel1923theory} suitable for metals. This discounts the applicability to 2D Dirac crystals of early models based on the polarizability of 2D gases in the context of short-range screening.\cite{maldague1978many}  

Because 2D Dirac crystals are zero-band gap semiconductors they dramatically amplify the effects of any phonon disturbance for which $q \approx 2k_\mathrm{F}$. This effect, known as Fermi-surface nesting (FSN) \cite{yan2020superconductivity,ali2016butterfly} and potentially leading to Kohn anomalies \cite{lazzeri2006nonadiabatic,kohn1959image} has been often considered using the Lindhard model. The Lindhard model is a method for calculating the effects of electric field screening by electrons in solids based on first-order quantum perturbation theory and it accurately predicts a common limitation of most of these calculations. For example, Kohn anomalies and FSN conditions for free electron gases of any dimensionality are correctly predicted using a static Lindhard model at 0 K\cite{mihaila2011lindhard}. In 2D Dirac crystals, the Lindhard model of electron screening by phonons has been considered by many authors such as \cite{wunsch2006dynamical,hwang2007dielectric,bahrami2017exchange,zhu2021dynamical,iurov2017exchange,lu2016friedel,calandra2007electron,lazzeri2006nonadiabatic}, however, because of the use of a $q$ independent from the disturbance energy $\hbar\omega_{0}$ in these reports none of them are well suited to estimate the Lindhard response function for acoustic phonon branches in 2D Dirac crystals. This is because a common limitation of most Lindhard model calculations is the use of the random phase approximation (RPA) in which the contribution to the dielectric response function from the total electric potential is assumed to average out so that only the potential at wave vector $k$ contributes. This considers only relatively weak electron screening potential and does not take into account the dispersion relationship $\omega_q$ of the phonon mode for which it is calculated, and cannot simply be obtained from replacing $\omega_q$ into $\chi(q,\omega)$.

Acoustic phonons, for which the energy $\hbar\omega_q=c\cdot q$ is proportional to the wavenumber via $c$, which is the speed of sound written in energy units are present in any crystalline lattice regardless of the number of atoms within their basis. Because acoustic phonons are responsible for the long-wavelength, low-energy end of the vibrational density-of-states, they are critically important for a host of measurable properties including but not limited to thermal transport \cite{hao2011mechanical,mahdizadeh2014thermal}, electric conductivity\cite{kozikov2010electron,kim2016electronic}, and Brillouin scattering \cite{wang2008brillouin,cong2019probing}, particularly at low or room temperatures where the optical phonons are not excited. Such properties are often calculated in 2D Dirac crystals by considering weak electron-phonon interaction and, consequently, short-range charge screening effects at the level of Thomas-Fermi.\cite{samaddar2016charge,hwang2007carrier} This is unreliable in the very frequent case in which small Fermi-energy fluctuations by doping or impurities produce Fermi level shifts bringing $k_\mathrm{F}$ within the range of acoustic phonon wavenumbers $q$. This may lead to FSN and anomalous electron-phonon interaction. For example, Ramezanali et al\cite{ramezanali2009finite} found that the accuracy of specific-heat calculations in graphene can be remarkably improved by introducing a Lindhard-based correction still assuming random-phase approximation and relatively weak electron-phonon coupling, which limits the generality of their approach. Furthermore, by deriving the dielectric response function as a function of $q$ using the Lindhard model we can improve on the phonon self-energy calculation performed in 2D Dirac crystals. The phonon line width derived from the phonon self-energy provides a way to gain experimental information about the electron-phonon coupling strength. The calculations for the phonon self-energy have been mostly carried out at $\mathrm{T}\to0$\cite{allen1972neutron} and $q=0$\cite{garcia2013coupling} which limits the application of such results. 

The objective of our work is to analytically calculate the dielectric response function of electrons strongly correlated with acoustic phonons by using the Lindhard model beyond the RPA in 2D Dirac crystals. Using scaling laws and introducing reduced phonon and electron wavevectors, respectively $\boldsymbol{\psi} $ and $\boldsymbol{\xi}$, as well as a reduced temperature $\tau$ and reduced speeds of sound $\mathrm{w_F}$  
\begin{equation}
    \boldsymbol{\xi}=\frac{\textbf{k}}{2k_\mathrm{{F}}}
    \qquad
    \boldsymbol{\psi}=\frac{\textbf{q}}{2k_\mathrm{{F}}}
    \qquad
    \tau=\frac{k_{B}\mathrm{T}}{2\mathrm{v_F}k_\mathrm{{F}}}
    \qquad
    \mathrm{w_F}=\frac{c}{\mathrm{v_F}},
\label{Defiening the dimensionless variables}
\end{equation}
(where $\mathrm{v_F} = E_\mathrm{F}/k_\mathrm{F}$ is the Fermi velocity) we show that a universal scaling law for the Lindhard dielectric response of acoustic phonons only depending on the dimensionless quantities in Eq.(\ref{Defiening the dimensionless variables}) can be established. Such an expression will be general enough to describe the electron-lattice interaction at any Fermi-level shifts and temperatures even for cases where the small-$\tau$ Sommerfeld approximation \cite{ashcroft2001festkorperphysik} customarily used in solid state physics is not valid. Further, using the derived dielectric response function we calculate the phonon line width from the phonon self-energy and compare our results with experimental data presented in the literature.

\section{\label{sec:level2}Methodology}
\begin{figure}[H]
\centering
\includegraphics[width=.75\columnwidth]{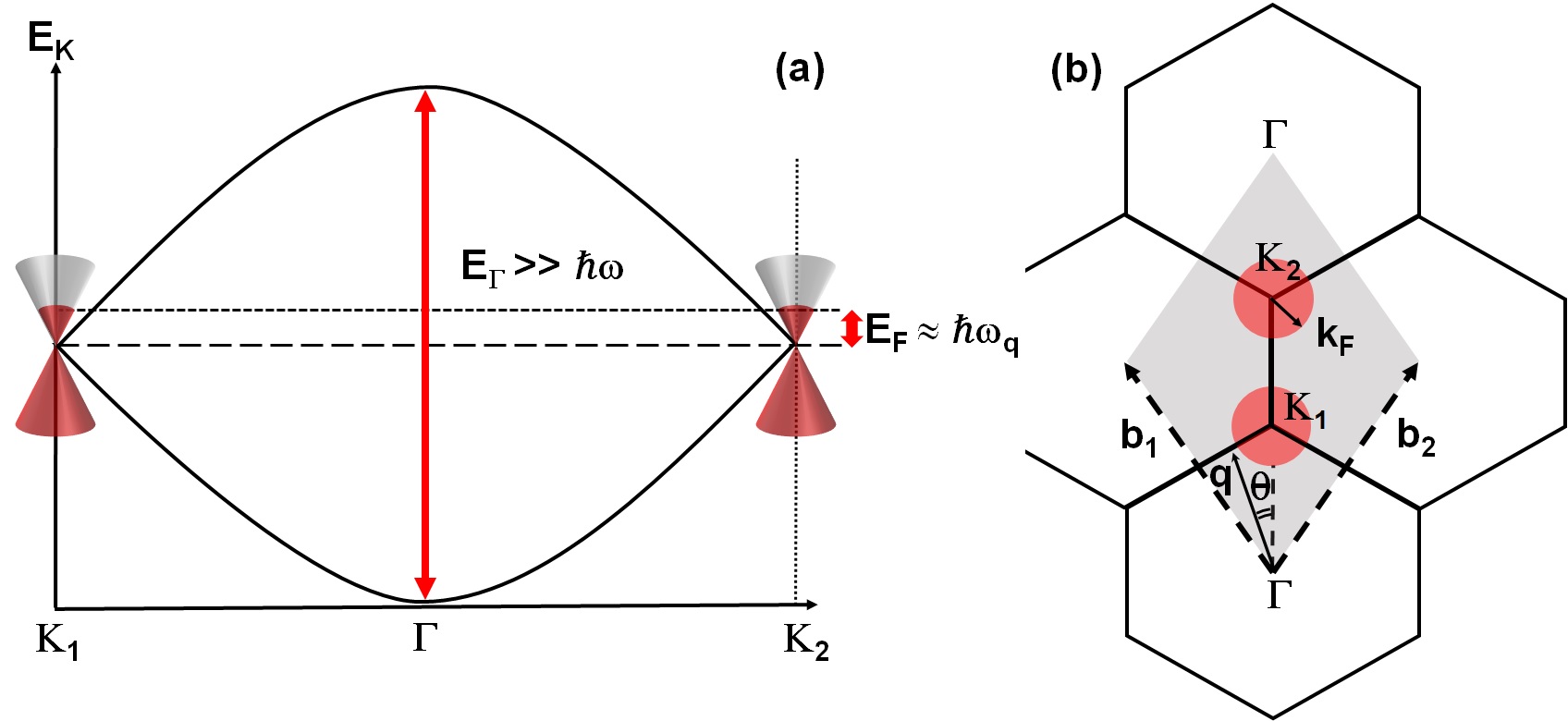}
\caption{(a) The $\pi$-$\pi*$ electron energy spacing along a straight line from the center to the bounds of a 2D Dirac lattice. The electron energy spacing at the zone center is much larger than the phonon energy making the electron-phonon interactions impossible. The electron energy spacing decreases and becomes comparable to the phonon energy as we move toward the lattice zone boundaries making the electron-phonon interactions possible. (b) The reciprocal lattice of a honeycomb structured lattice is shown as the grey-shaded region where we have two Dirac cones in which electron-phonon interactions take place at.}
\label{Figure1}
\end{figure}

\begin{figure}[H]
\centering
\includegraphics[width=.75\columnwidth]{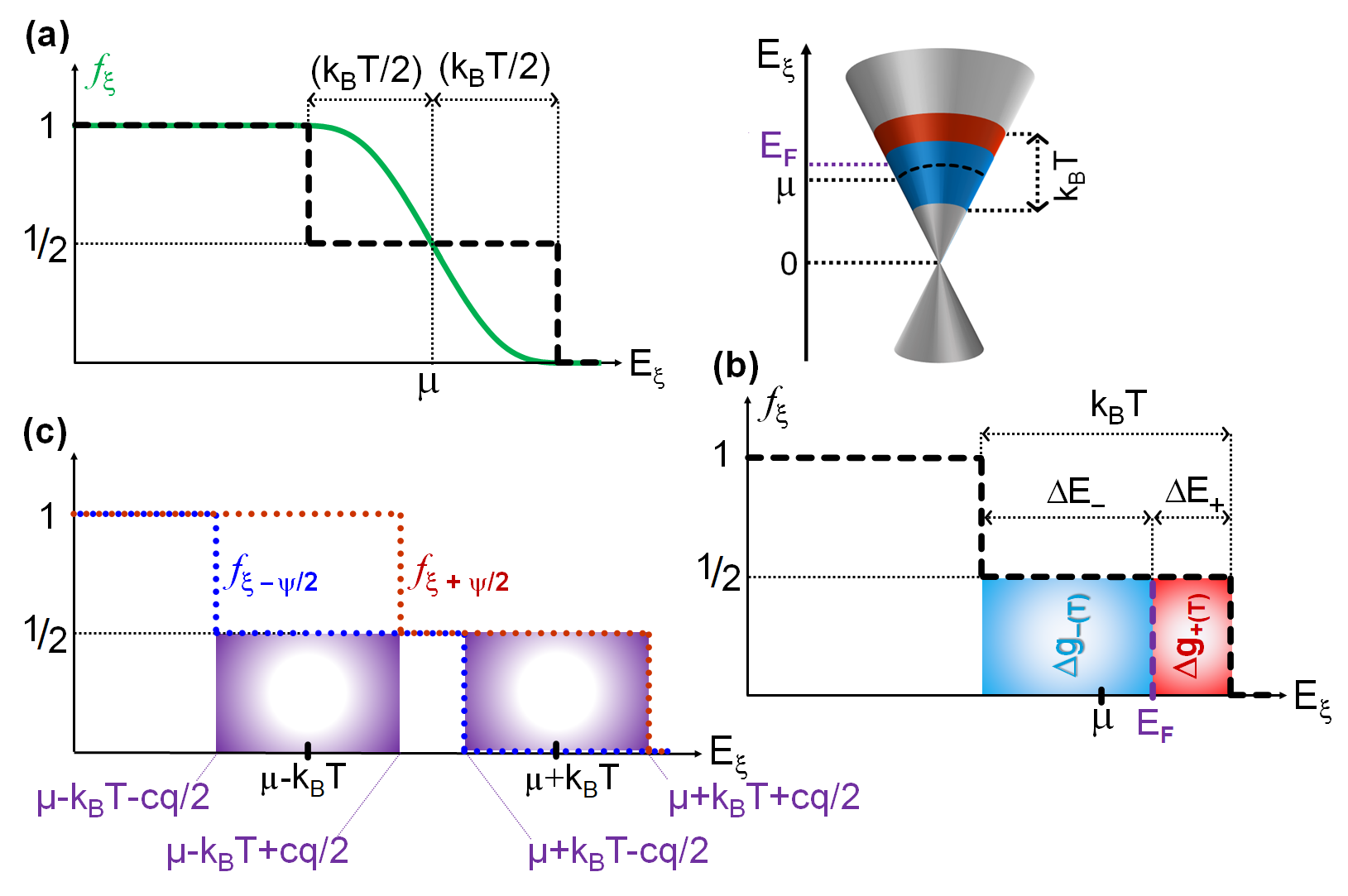}
\caption{(a) “Double step” approximation of the occupation function of an electron used in this paper to calculate the dielectric response function for $\mathrm{T}\ne0$ K. (b) The “double step” approximation and the Dirac cone of a 2D crystal with an elevated Fermi energy level $E_{\mathrm{F}}$. Areas of the Dirac cone highlighted in red and blue lead to equal number of states at energy ranges $\Delta E_{+}$ and $\Delta E_{-}$ above and below $E_{\mathrm{F}}$, respectively. (c) The difference between $ f_{\boldsymbol{\xi}+\psi/2}$ and $f_{\boldsymbol{\xi}-\psi/2}$ as a function of energy in a 2D Dirac crystal shown in the two purple squares which set the bounds of integration for calculating the dielectric response function.}
\label{Figure2}
\end{figure}

Using the normalized quantities in Eq.(\ref{Defiening the dimensionless variables}) the Lindhard dielectric response function\cite{ashcroft2001festkorperphysik,maldague1978many} takes the form: 
\begin{equation}
    \chi(q,\omega)=-\frac{(2{k}_{\mathrm{F}})e^{2}}{2\pi^2} \int \frac{f_{\boldsymbol{\xi}-\frac{\psi}{2}}-f_{\boldsymbol{\xi}+\frac{\psi}{2}}} {E_{\boldsymbol{\xi}+\frac{\psi}{2}}-{E_{\boldsymbol{\xi}-\frac{\psi}{2}}}-\hbar\omega_{\psi}} d\boldsymbol{\xi},
\label{main equation for dielectric response function}
\end{equation}
where $f_{\boldsymbol{\xi}\pm\psi/{2}}$ are the occupation function of the single particle state following the Fermi-Dirac distribution and $\hbar\omega_{\psi}$ is the phonon energy. The integral is over the momentum space of the Fermi sphere and $E_{\boldsymbol{\xi}\pm\psi/2}$ are the energy of the created and annihilated electrons, respectively. We do not integrate over the electron spins and have simplified the problem by multiplying Eq.(\ref{main equation for dielectric response function}) by a factor of 2. As shown in Fig.~\ref{Figure1}(a) the $\pi$-$\pi*$ electron energy spacing in a 2D Dirac crystal at the zone center $E_{\Gamma}$\cite{geim2010rise,katsnelson2007graphene} is too high relative to the energy of acoustical phonons \cite{alofi2014theory} $E_{ph} = \hbar\omega_{\psi}$ for any electron-phonon interactions to take place without violating the conservation of energy. The electron energy spacing becomes comparable to the energy of phonons at the \textbf{K}-point thus making the electron-phonon interactions possible. Assuming a honeycomb lattice structure as shown in  Fig.~\ref{Figure1}(b) the electron-phonon interactions in the reciprocal lattice shown by the shaded grey region occur at the two-zone boundary points K1 and K2 depicted by the Dirac cones. 
To derive an analytical expression for the Lindhard dielectric response function we must solve the difference between the occupation number of the levels above and below the Fermi energy, which is provided by the Fermi Dirac statistics in the numerator of the integral in Eq.(\ref{main equation for dielectric response function}) by applying a suitable approximation. The most used approximation for simplifying the Fermi-Dirac distribution function is the “single step” function used at $\mathrm{T}\approx 0$ K where the Fermi-Dirac distribution has a value of 1 for energies below the Fermi energy, and a value of 0 for energies above.\cite{pathria2016statistical} However, at finite temperatures the “single step” function approximation is no longer accurate since the distribution gets smeared out, as some electrons begin to be thermally excited to energy levels above $E_{\mathrm{F}}$. We, therefore, approximate the Fermi-Dirac distribution with a “double step” function which has three values 1, 1/2, and 0 and covers the distribution of electrons at different energies more accurately, shown in Fig.~\ref{Figure2}(a). Furthermore, while at T = 0 K the chemical potential $\mu$ is equal to $E_{\mathrm{F}}$ this is no longer the case at finite temperatures where $\mu$ becomes dependent on both $E_{\mathrm{F}}$ and T. Therefore, to find the values of the Fermi-Dirac distribution function at different energies using the “double step” function we must find a relation between $\mu$, $E_{\mathrm{F}}$ and T. 

In metals where we have large free electron density in the proximity of the Fermi energy $E_{\mathrm{F}}$, we use the Sommerfeld expansion to write the chemical potential, $\mu$, as a function of $E_{\mathrm{F}}$ and T \cite{ashcroft2001festkorperphysik}. However, Sommerfeld expansion cannot be reliable for 2D Dirac crystals which have a limited concentration of free electrons and we need to express small fluctuations of $\mu$ over $E_{\mathrm{F}}$. Therefore, to write $\mu$ in terms of $E_{\mathrm{F}}$ and T for 2D Dirac crystals we have to develop a new relation. To this end, we define $\Delta E_{\pm}$ to be the range of the singly occupied energy levels in the Dirac cone situated above and below the Fermi energy, respectively, shown in Fig.~\ref{Figure2}(b). The integrated density of states over $\Delta E_{\pm}$ are equal and are defined as $\Delta g_{\pm \mathrm{(T)}}$. By analyzing Fig.~\ref{Figure2}(b) we can write $\mu$ as a function of $E_{\mathrm{F}}$ and $\Delta E_{\pm}$ in the following manner: 
\begin{equation}
    \mu=E_{F}+\frac{\Delta E_{+}-\Delta E_{-}}{2}.
\label{The chemical potantial in terms of E_F and Delta(E_+) and Delta(E_-)}
\end{equation}
To write $\Delta E_{\pm}$ in terms of $E_{\mathrm{F}}$ and T we first write the sum of the singly occupied energy levels in the Dirac cone situated above and below the Fermi energy level, Fig.~\ref{Figure2}(b), as follow:
\begin{equation}
    \Delta E_{+}+\Delta E_{-}=k_{B}\mathrm{T}.
\label{The sum of Delta(E_+) and Delta(E_-)}
\end{equation}
We further know that the integrated density of states over $\Delta E_{+}$ and $\Delta E_{-}$ are equal. This results in the surface area of the two red and blue regions on the Dirac cone in Fig.~\ref{Figure2}(b) to be equal to one another and result in the following equation:
\begin{equation}
    \Delta E_{-}(2E_{\mathrm{F}}-\Delta E_{-})=\Delta E_{+}(2E_{\mathrm{F}}+\Delta E_{+}).
\label{Putting the two integrated density of states equal}
\end{equation}
By inserting Eq.(\ref{The sum of Delta(E_+) and Delta(E_-)}) into Eq.(\ref{Putting the two integrated density of states equal}) we find a relation between $\Delta E_{\pm}$ and $E_{\mathrm{F}}$ and T. We can therefore write the chemical potential as:
\begin{equation}
    \mu(E_{\mathrm{F}},\mathrm{T})=E_{\mathrm{F}}+\frac{\Delta E_{+}-\Delta E_{-}}{2}=\sqrt{\Big|E_{\mathrm{F}}^{2}-\Big(\frac{k_B \mathrm{T}}{2}\Big)^{2}\Big|}.
\label{Writting mu in terms of E_F and T}
\end{equation}
To find the difference between $f_{\boldsymbol{\xi}+\psi/{2}}$ and $f_{\boldsymbol{\xi}-\psi/{2}}$ versus the energy we use a “double step” function approximation as shown in Fig.~\ref{Figure2}(a). By computing the area under the curve for the two occupation levels shown in Fig.~\ref{Figure2}(c) we have:
\begin{equation}
    f_{\boldsymbol{\xi}+\frac{\psi}{2}}-f_{\boldsymbol{\xi}-\frac{\psi}{2}} =
\left\{
\begin{array}{ll}
		``1/2"  & \sqrt{\Big|E_{\mathrm{F}}^{2}-\Big(\frac{k_B \mathrm{T}}{2}\Big)^{2}\Big|}-\frac{k_B\mathrm{T}}{2}-\frac{cq}{2}<E_{\xi}<\sqrt{\Big|E_{\mathrm{F}}^{2}-\Big(\frac{k_B \mathrm{T}}{2}\Big)^{2}\Big|}-\frac{k_B \mathrm{T}}{2}+\frac{cq}{2} \\
		``1/2" &  \sqrt{\Big|E_{\mathrm{F}}^{2}-\Big(\frac{k_B \mathrm{T}}{2}\Big)^{2}\Big|}+\frac{k_B\mathrm{T}}{2}-\frac{cq}{2}<E_{\xi}<\sqrt{\Big|E_{\mathrm{F}}^{2}-\Big(\frac{k_B \mathrm{T}}{2}\Big)^{2}\Big|}+\frac{k_B \mathrm{T}}{2}+\frac{cq}{2}
	\end{array}
\right.
\label{Difference between the two Fermi-Dirac distributions}
\end{equation}
Where the energy bounds in Eq.(\ref{Difference between the two Fermi-Dirac distributions}) for which $f_{\boldsymbol{\xi}+\psi/{2}}-f_{\boldsymbol{\xi}-\psi/{2}}=1/2$ is shown by the purple shaded region in Fig.~\ref{Figure2}(c). The difference between the two, sets the bounds of integration for calculating the dielectric response function at non-zero temperature. By analyzing Eq.(\ref{Difference between the two Fermi-Dirac distributions}) at $\mathrm{T}\approx0$ K we observe that the two purple regions in Fig.~\ref{Figure2}(c) overlap with one another, and the “double step” function turns into a “single step” function further confirming our results.

By raising the temperature, the Fermi energy of the 2D Dirac crystal shifts leading the bounds of the integration in Eq.(\ref{main equation for dielectric response function}) for the dielectric response function to also change. While the bounds of the integration for $\xi(\psi,\omega)$ at T = 0 K are between $1/2\pm\psi/2$, using the “double step” function to derive the difference in the Fermi-Dirac distribution for $\mathrm{T}\ne0$ K we have: 
\begin{equation}
\begin{split}
   \chi(\psi &,\omega)=-\frac{e^{2}}{4\pi^2}\\ &\Bigg( \int_{\frac{\mu}{2E_{\mathrm{F}}}-\frac{\tau}{2}-\frac{\psi}{2}}^{\frac{\mu}{2E_{\mathrm{F}}}-\frac{\tau}{2}+\frac{\psi}{2}} d\xi_x \int_{\xi_{y,min}}^{\xi_{y,max}} \frac{d\xi_y}{E_{\boldsymbol{\xi}+\frac{\psi}{2}}-{E_{\boldsymbol{\xi}-\frac{\psi}{2}}}-cq}+\int_{\frac{\mu}{2E_{\mathrm{F}}}+\frac{\tau}{2}-\frac{\psi}{2}}^{\frac{\mu}{2E_{\mathrm{F}}}+\frac{\tau}{2}+\frac{\psi}{2}} d\xi_x \int_{\xi_{y,min}}^{\xi_{y,max}} \frac{d\xi_y}{E_{\boldsymbol{\xi}+\frac{\psi}{2}}-{E_{\boldsymbol{\xi}-\frac{\psi}{2}}}-cq}\Bigg).
\label{Dielectric susceptibility at nonzero temperature}
\end{split}
\end{equation}
Where we have written the dielectric response function in dimensionless coordinates as $\chi(\psi,\omega)=\chi(q,\omega)/2k_{\mathrm{F}}$. We see that by approximating the occupation function of the electron with a “double step” function we split the integral in Eq.(\ref{main equation for dielectric response function}) into two integrals where the bounds of each integral is one of the purple shaded regions depicted in  Fig.~\ref{Figure2}(c).

A disadvantage of the Lindhard model over more simplistic approximations such as Thomas-Fermi, is the requirement of carrying out integrations that often cannot be performed analytically, and may become cumbersome where: $i)$ the model is dynamic–i.e. one assumes that phonons not only carry momentum, but also energy $\hbar\omega_{0}$; or $ii)$ nonzero temperature is considered–a necessary requirement to compare $\chi(q)$ with experiments involving, for example, T-dependent electrical or transport measurements; or $iii)$ the phonon energy $\hbar\omega_{q}$ is wavevector-dependent in a dispersive system–a respect in which it is worthwhile noting that the phonon energy is not only required in dynamic Lindhard-model calculations but also static ones, as it affects the electron Fermi-Dirac distribution, also at 0 K. Of course, the complexity of the calculation will increase if more than one of the phenomena $(i-iii)$ are considered. To simplify the integral we assume the extreme carrier mobility in 2D Dirac crystals results in their Fermi velocity to be much larger than the speed of sound, $\mathrm{v}_{\mathrm{F}}>>c$. This has been experimentally confirmed in various 2D Dirac crystals such as graphene\cite{hwang2012fermi}, borophene\cite{xu2016hydrogenated}, Weyl semimetals\cite{lee2015fermi}, and silicene\cite{kara2012review} where $c < 0.01 \mathrm{v}_{\mathrm{F}}$. Defining $\mathrm{w}_{\mathrm{F}}=c/\mathrm{v}_{\mathrm{F}}$ we can expand the dielectric response function around $\mathrm{w}_{\mathrm{F}}$ with $\mathrm{w}_{\mathrm{F}}\to0$ in the following manner:
\begin{equation}
    \chi_{\mathrm{w_{F}}} (\psi)=\chi_{0} (\psi)+\pdv[]{\chi_{0} (\psi)}{\mathrm{w_{F}}}\mathrm{w_{F}}+O(w_{F}^{2}),
\label{Expanding the dielectric susceptibility in terms of w_F}
\end{equation}
The first addend in Eq.(\ref{Expanding the dielectric susceptibility in terms of w_F}) is the static dielectric response of the 2D Dirac crystal with the phonon energy being equal to zero. The next addends are the higher order terms which are the dielectric response at nonzero phonon energy. Due to computational convenience in the calculation, we shall derive the static term in the Cartesian coordinates and the dynamic terms in polar counterparts. In the end, by adding the static term with the dynamic terms we get the dynamic dielectric response function of a 2D Dirac crystal.
\begin{figure}[H]
\centering
\includegraphics[width=.45\columnwidth]{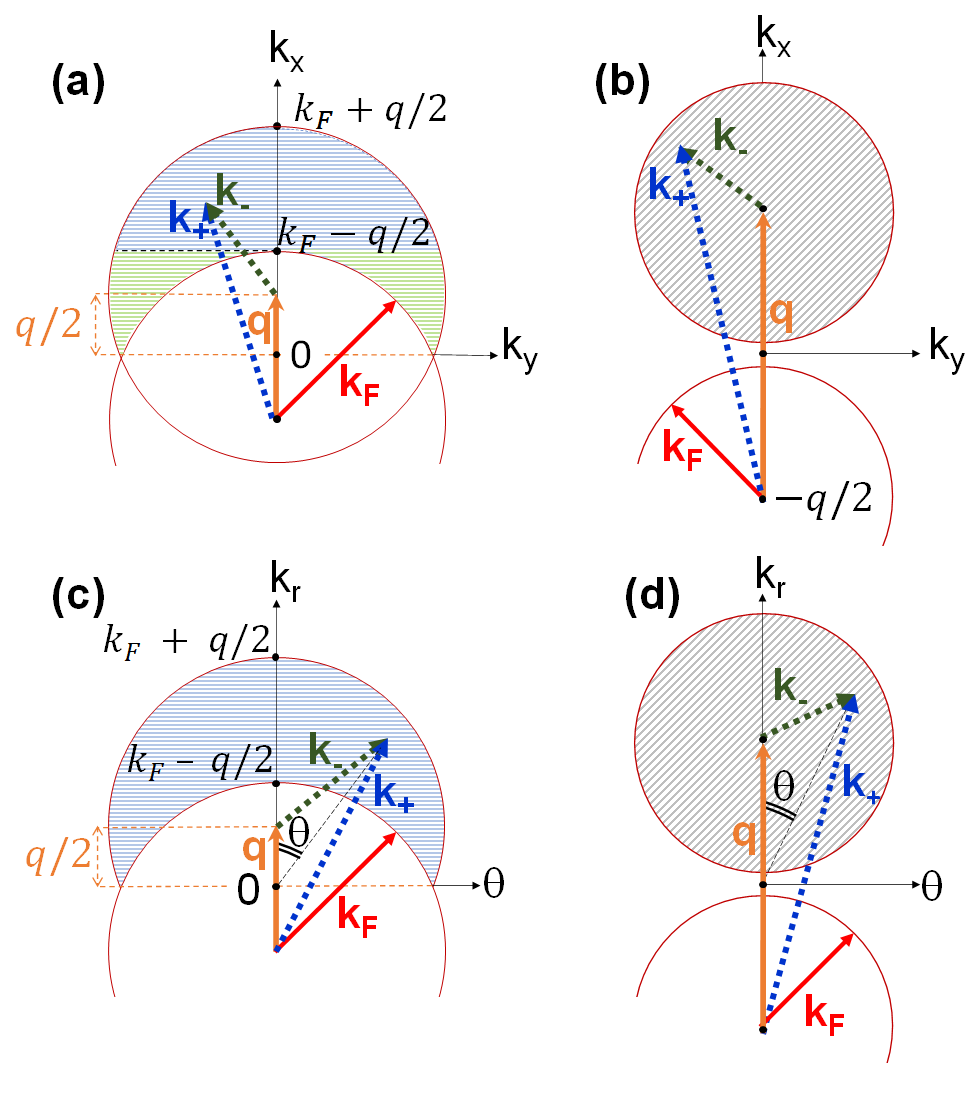}
\caption{(a),(b) The shift in the momentum sphere respectively for $q<2k_{\mathrm{F}}$ or $\psi<1$ and $q>2k_{\mathrm{F}}$ or $\psi>1$ written in the Cartesian coordinate system. We write the shift in the momentum sphere in Cartesian coordinates to calculate the static dielectric response function. (c), (d) The shift in the momentum sphere respectively for $q<2k_{\mathrm{F}}$ or $\psi<1$ and $q>2k_{\mathrm{F}}$ or $\psi>1$ written in the polar coordinate system. We write the shift in the momentum sphere in polar coordinates to calculate the 2nd to last addends of the dielectric response function which contain the phonon energy.}
\label{Figure3}
\end{figure}

\section{\label{sec:level3}Results}
To derive the dielectric response function, we assume we have a shift in the momentum sphere\cite{nakamura2020robust,lokot2014creation} equal to $q=(2k_{\mathrm{F}})\psi$, caused by a disturbance in the lattice. Fig.~\ref{Figure3} reports the shift in the momentum sphere written in the Cartesian and polar coordinate systems. Two different trends for the dielectric response function are obtained in the two regions at $q<2k_{\mathrm{F}}$ or $\psi<1$ and $q>2k_{\mathrm{F}}$ or $\psi>1$, respectively.

\subsection{\label{sec:level3-1}Static dielectric response}

The integral in Eq.(\ref{The initial first term of the dielectric susceptibility}) expresses the static dielectric response, $\chi_{0}(\psi)$, function of a 2D Dirac beyond the RPA in the two regions of $\psi<1$ and $\psi>1$. Following Eq.(\ref{main equation for dielectric response function}) $\chi_{0}(\psi)$ of a 2D Dirac crystal at T = 0 K is equal to:
\begin{equation}
\begin{split}
    &\chi _{0}(\psi)=-\frac{e^{2}}{2\pi^2 \mathrm{v}_\mathrm{F}} \\&\Bigg( \int_{0}^{\frac{1}{2}-\frac{\psi}{2}}d\xi{_x}
    \int_{\frac{1}{2}\left[1-(\xi_{x}+\frac{\psi}{2})^{2}\right]^{1/2}}^ 
    {\frac{1}{2}\left[1+(\xi_{x}+\frac{\psi}{2})^{2}\right]^{1/2}} \frac{d\xi_y}{\sqrt{(\xi_{x}+\psi/2)^{2}+\xi_y^{2}}-\sqrt{(\xi_{x}-\psi/2)^{2}+\xi_y^{2}}} \\& +\int_{\frac{1}{2}-\frac{\psi}{2}}^{\frac{1}{2}+\frac{\psi}{2}}d\xi_x
    \int_{0}^{\left[1-(\xi_{x}-\frac{\psi}{2})^{2}\right]^{1/2}} \frac{d\xi_y}{\sqrt{(\xi_{x}+\psi/2)^{2}+\xi_y^{2}}-\sqrt{(\xi_{x}-\psi/2)^{2}+\xi_y^{2}}} \\& +\int_{\frac{\psi}{2}-\frac{1}{2}}^{\frac{\psi}{2}+\frac{1}{2}}d\xi_x
    \int_{0}^{\frac{1}{2}\left[1+(\xi_{x}-\frac{\psi}{2})^{2}\right]^{1/2}} \frac{d\xi_y}{\sqrt{(\xi_{x}+\psi/2)^{2}+\xi_y^{2}}-\sqrt{(\xi_{x}-\psi/2)^{2}+\xi_y^{2}}}
    \Bigg).
\label{The initial first term of the dielectric susceptibility}
\end{split}
\end{equation}
Since we integrate over the two momentum spheres Fig.~\ref{Figure3}, we multiply $\chi_{0}(\psi)$ in Eq.(\ref{main equation for dielectric response function}) by a factor of 2. Furthermore, since the bounds along the y-axis change when writing the integral in the Cartesian coordinate system for $\psi<1$, we divide the integral into two, with each integral covering a separate region as shown in Fig.~\ref{Figure3}(a) by the two blue and green colors. So, the first two integrals in Eq.(\ref{The initial first term of the dielectric susceptibility}) are for the region $\psi<1$ and the third integral is for the region $\psi>1$. Analyzing each of these two regions separately, for $\psi<1$ we have: 
\begin{equation}
\begin{split}
    -\bigg(&\frac{2\pi^2 \mathrm{v}_\mathrm{F}}{e^{2}}\bigg)\chi _{0}(\psi)= \frac{1}{\psi} \lim_{\epsilon \to 0} \Bigg(\int_{0}^{\frac{1+\psi-\epsilon}{2}}d\xi \Bigg[\frac{\sqrt{1-(2\xi-\psi)^{2}}}{16\xi}\Big(1+\sqrt{1+8\psi\xi}\Big)+\\& \frac{(2\xi+\psi)^{2}}{8\xi} \asinh\bigg({\frac{\sqrt{1-(2\xi-\psi)^{2}}}{2\xi+\psi}}\bigg)+\frac{(2\xi-\psi)^{2}}{8\xi} \asinh\bigg({\frac{\sqrt{1-(2\xi-\psi)^{2}}}{2\xi-\psi}}\bigg)\Bigg]\\& -\int_{0}^{\frac{1-\psi-\epsilon}{2}}d\xi\Bigg[\frac{\sqrt{1-(2\xi+\psi)^{2}}}{16\xi}\Big(1+\sqrt{1-8\psi\xi}\Big)+\\& \frac{(2\xi+\psi)^{2}}{8\xi} \asinh\bigg({\frac{\sqrt{1-(2\xi+\psi)^{2}}}{2\xi+\psi}}\bigg)+\frac{(2\xi-\psi)^{2}}{8\xi} \asinh\bigg({\frac{\sqrt{1-(2\xi+\psi)^{2}}}{2\xi-\psi}}\bigg)\Bigg]\Bigg).
\label{Taking the first term of the dielectric susceptibility to dimensionless coordinates for psi < 1}
\end{split}
\end{equation}
where we have integrated over $\xi_y$. The employment of numerical methods in different branches of physics and science has become more common in recent years \cite{ezugwu2017contactless,afrasiabian2023dispersion,farhani2023bayesian,farhani2021classification,yazdi2022nemo}. Using a MATLAB routine, we solve Eq.(\ref{Taking the first term of the dielectric susceptibility to dimensionless coordinates for psi < 1}) numerically by sums. The designated solid black line in Fig.~\ref{Figure4}(a) numbered (\textbf{I}) is the integral solution of Eq.(\ref{Taking the first term of the dielectric susceptibility to dimensionless coordinates for psi < 1}). To further analyze the integral and come up with an analytical solution we partition the different parts of Eq.(\ref{Taking the first term of the dielectric susceptibility to dimensionless coordinates for psi < 1}). The red dashed lines numbered ($i$) and ($ii$) are the two positive and two negative terms containing the arc hyperbolic sinus. The sum of these two terms is ($i$) + ($ii$) = $(\textbf{I}_{3})$ shown in Fig.~\ref{Figure4}(a) as a solid red line. 
\begin{equation}
\begin{split}
    &(\textbf{I}_{3})=(i)+(ii)=\frac{1}{\psi}  \lim_{\epsilon \to 0} \\&\Bigg(\int_{0}^{\frac{1+\psi-\epsilon}{2}}d\xi\Bigg[ \frac{(2\xi+\psi)^{2}}{8\xi} \asinh\bigg({\frac{\sqrt{1-(2\xi-\psi)^{2}}}{2\xi+\psi}}\bigg)+\frac{(2\xi-\psi)^{2}}{8\xi} \asinh\bigg({\frac{\sqrt{1-(\xi-\psi)^{2}}}{2\xi-\psi}}\bigg)\Bigg] -\\&\int_{0}^{\frac{1-\psi-\epsilon}{2}}d\xi\Bigg[ \frac{(2\xi+\psi)^{2}}{8\xi} \asinh\bigg({\frac{\sqrt{1-(2\xi+\psi)^{2}}}{2\xi+\psi}}\bigg)+\frac{(2\xi-\psi)^{2}}{8\xi} \asinh\bigg({\frac{\sqrt{1-(\xi+\psi)^{2}}}{2\xi-\psi}}\bigg)\Bigg]\Bigg)\\&\approx 0.39.
\label{The sum of the arcsinh terms of the static dielectric susceptibility for psi < 1}
\end{split}
\end{equation}
It can be seen in Fig.~\ref{Figure4}(a) that the designated line $(\textbf{I}_{3})$ is approximately a constant equal to $(\textbf{I}_{3})\approx0.39$. This can be proven by Taylor expanding $(\textbf{I}_{3})$ up to the first order about $\psi = 0$. We further separate the remaining terms in Eq.(\ref{Taking the first term of the dielectric susceptibility to dimensionless coordinates for psi < 1}) in the following manner:
\begin{equation}
\begin{split}
    (\textbf{I}_{1})&=\frac{1}{\psi} \lim_{\epsilon \to 0}  \Bigg(\int_{0}^{\frac{1+\psi-\epsilon}{2}}d\xi\Bigg[\frac{\sqrt{1-(2\xi-\psi)^{2}}}{16\xi}- \int_{0}^{\frac{1-\psi-\epsilon}{2}}d\xi\Bigg[\frac{\sqrt{1-(2\xi+\psi)^{2}}}{16\xi}\Bigg]\Bigg)\\ & = \frac{\pi\psi+\sqrt{1-\psi^{2}} \ln{\Big(\frac{1-\psi}{1+\psi}}\Big)} {16\psi},
\label{The sum of the analytical terms of the static dielectric susceptibility for psi < 1}
\end{split}
\end{equation}
\begin{equation}
\begin{split}
    (\textbf{I}_{2})=\frac{1}{\psi} \lim_{\epsilon \to 0}  \Bigg(&\int_{0}^{\frac{1+\psi-\epsilon}{2}}d\xi\Bigg[\frac{\sqrt{1-(2\xi-\psi)^{2}}}{16\xi}\Big(\sqrt{1+8\psi\xi}\Big)- \\ &\int_{0}^{\frac{1-\psi-\epsilon}{2}}d\xi\Bigg[\frac{\sqrt{1-(2\xi+\psi)^{2}}}{16\xi}\Big(\sqrt{1-8\psi\xi}\Big)\Bigg]\Bigg).
\label{The sum of the non analytical terms of the static dielectric susceptibility for psi < 1}
\end{split}
\end{equation}
Where (\textbf{I}$_1$) is shown in a purple solid line and (\textbf{I}$_2$) is shown in a blue solid line in Fig.~\ref{Figure4}(a). By analyzing Fig.~\ref{Figure4}(a) we observe that (\textbf{I}$_2$) can be estimated as (\textbf{I}$_2$) $\approx$ 0.2 + (\textbf{I}$_1$) / (2$\psi$). This can also be proven by Taylor expanding (\textbf{I}$_1$) and (\textbf{I}$_2$) up to the second order about the midpoint of the integral bounds which are (1 + $\psi$)/4 and (1 - $\psi$)/4. We can therefore write Eq. \ref{Taking the first term of the dielectric susceptibility to dimensionless coordinates for psi < 1} as:
\begin{equation}
\begin{split}
    -\bigg(&\frac{2\pi^2 \mathrm{v}_\mathrm{F}}{e^{2}}\bigg)\chi _{0}(\psi)= (\textbf{I}_1)+(\textbf{I}_2)+(\textbf{I}_3)= 0.59+\frac{3\Big[\pi\psi+\sqrt{1-\psi^{2}} \ln{\Big(\frac{1-\psi}{1+\psi}}\Big)\Big]} {32\psi} .
\label{The static dielectric susceptibility in dimensionless coordinates for psi < 1}
\end{split}
\end{equation}
Where Eq.(\ref{The static dielectric susceptibility in dimensionless coordinates for psi < 1}) is the analytical expression of the static dielectric response of a 2D Driac crystal in the region of $\psi<1$. For $\psi>1$ we have:
\begin{equation}
\begin{split}
    -\bigg(&\frac{2\pi^2 \mathrm{v}_\mathrm{F}}{e^{2}}\bigg)\chi _{0}(\psi)= \frac{1}{\psi}\Bigg( \int_{\frac{1-\psi}{2}}^{\frac{1+\psi}{2}}d\xi \Bigg[\frac{\sqrt{1-(2\xi-\psi)^{2}}}{16\xi}\Big(1+\sqrt{1+8\psi\xi}\Big)+\\& \frac{(2\xi+\psi)^{2}}{8\xi} \asinh\bigg({\frac{\sqrt{1-(2\xi-\psi)^{2}}}{2\xi+\psi}}\bigg)+\frac{(2\xi-\psi)^{2}}{8\xi} \asinh\bigg({\frac{\sqrt{1-(2\xi-\psi)^{2}}}{2\xi-\psi}}\bigg)\Bigg]\Bigg).
\label{The static term of the dielectric susceptibility in dimensionless coordinates for psi > 1}
\end{split}
\end{equation}
The designated solid black line numbered (\textbf{I$^\prime$}) in Fig.~\ref{Figure4}(a) is the integral solution of Eq.(\ref{The static term of the dielectric susceptibility in dimensionless coordinates for psi > 1}) solved numerically. We again partition the different parts of the integral to come up with an analytical solution. The red dashed lines in Fig.~\ref{Figure4}(a) numbered ($i^\prime$) and ($ii^\prime$) are the first and second terms containing the arc hyperbolic sin term in Eq.(\ref{The static term of the dielectric susceptibility in dimensionless coordinates for psi > 1}). The sum of these two terms is ($i^\prime$) + ($ii^\prime$) = (\textbf{I$_3 ^\prime$}), shown in Fig.~\ref{Figure4}(a) as a solid red line.
\begin{equation}
\begin{split}
    &(\textbf{I}^{\prime}_{3})=(i^{\prime})+(ii^{\prime})= \\&\frac{1}{\psi} \Bigg(\int_{\frac{1-\psi}{2}}^{\frac{1+\psi}{2}}d\xi    \Bigg[\frac{(2\xi+\psi)^{2}}{8\xi} \asinh\bigg({\frac{\sqrt{1-(2\xi-\psi)^{2}}}{2\xi+\psi}}\bigg)+\frac{(2\xi-\psi)^{2}}{8\xi} \asinh\bigg({\frac{\sqrt{1-(\xi-\psi)^{2}}}{2\xi-\psi}}\bigg)\Bigg]\Bigg)\\&\approx \frac{0.39}{\psi}.
\label{The sum of the arcsinh terms of the static dielectric susceptibility for psi < 1}
\end{split}
\end{equation}
This can be proven by Taylor expanding (\textbf{I$_3 ^\prime$}) up to the first order about the midpoint of the integral bound which is $\psi/2$. We further separate the remaining terms in Eq. \ref{The static term of the dielectric susceptibility in dimensionless coordinates for psi > 1} as following:
\begin{equation}
\begin{split}
    (\textbf{I}^{\prime}_{1})&=\frac{1}{\psi} \Bigg(\int_{\frac{1-\psi}{2}}^{\frac{1+\psi}{2}}d\xi\Bigg[\frac{\sqrt{1-(2\xi-\psi)^{2}}}{16\xi}\Bigg]\Bigg) = \frac{\pi+\sqrt{1-\psi^{2}} \ln{\Big(\frac{1-\psi}{-1+\psi}}\Big)} {16\psi},
\label{The sum of the analytical terms of the static dielectric susceptibility for psi > 1}
\end{split}
\end{equation}
\begin{equation}
\begin{split}
    (\textbf{I}^{\prime}_{2})=\frac{1}{\psi} \Bigg(&\int_{\frac{1-\psi}{2}}^{\frac{1+\psi}{2}}d\xi\Bigg[\frac{\sqrt{1-(2\xi-\psi)^{2}}}{16\xi}\Big(\sqrt{1+8\psi\xi}\Big)\Bigg]\Bigg).
\label{The sum of the non analytical terms of the static dielectric susceptibility for psi > 1}
\end{split}
\end{equation}
Where (\textbf{I$_1 ^\prime$}) is shown by a purple solid line while (\textbf{I$_2 ^\prime$}) is shown by a blue solid line in Fig.~\ref{Figure4}(a). By analyzing Fig.~\ref{Figure4}(a) we observe that (\textbf{I$_2 ^\prime$}) can be estimated as (\textbf{I$_2 ^\prime$}) $\approx$ [0.4 + (\textbf{I$_1 ^\prime$})] / (2$\psi$). This can also be proven by Taylor expanding (\textbf{I$_1 ^\prime$}) and (\textbf{I$_2 ^\prime$}) up to the first order about their midpoint integral bound which is $\psi$/2. We can therefore write Eq.(\ref{The static term of the dielectric susceptibility in dimensionless coordinates for psi > 1}) as:
\begin{equation}
\begin{split}
    -\bigg(&\frac{2\pi^2 \mathrm{v}_\mathrm{F}}{e^{2}}\bigg)\chi _{0}(\psi)= (\textbf{I}^{\prime}_1)+(\textbf{I}^{\prime}_2)+(\textbf{I}^{\prime}_3)= \frac{1}{\psi}\Bigg(0.59 + \frac{3\Big[\pi\psi+\sqrt{1-\psi^{2}} \ln{\Big(\frac{1-\psi}{1+\psi}}\Big)\Big]}{32}\Bigg) .
\label{The static dielectric susceptibility in dimensionless coordinates for psi > 1}
\end{split}
\end{equation}
Where Eq.(\ref{The static dielectric susceptibility in dimensionless coordinates for psi > 1}) is the analytical expression of the static dielectric function of a 2D Driac crystal in the region of $\psi>1$. The comparison between the analytical and numerical solution of the static dielectric response using the Lindhard model is shown in Fig.~\ref{Figure4}(b) in which the two solutions are in good agreement with one another. By studying both graphs we see their cuspidal points are at $\psi= 1$ or $q = 2k_{F}$ which is the Fermi wave number. We, therefore, have the absolute value of the static dielectric response function increase up to $\psi = 1$ and then decrease as we go further away. 
\subsection{\label{sec:level3-2}	Dielectric response at nonzero phonon energy}
\begin{figure}[h]
\centering
\includegraphics[width=.75\columnwidth]{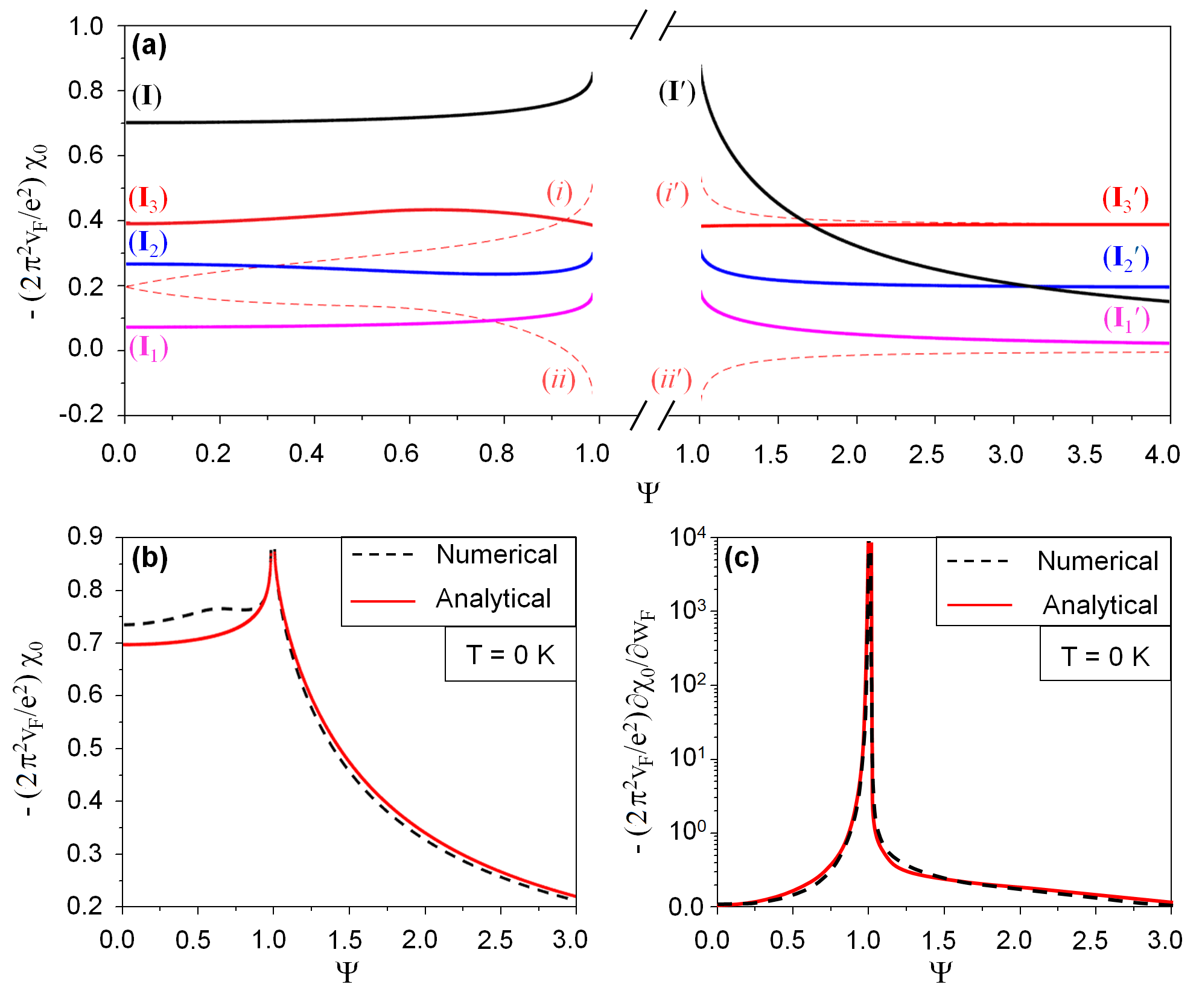}
\caption{(a) The dashed red lines on the left and the red dashed lines on the right are the terms containing the arc hyperbolic sinus terms in Eq.(\ref{Taking the first term of the dielectric susceptibility to dimensionless coordinates for psi < 1}) and Eq.(\ref{The static term of the dielectric susceptibility in dimensionless coordinates for psi > 1}), respectively. These terms are solved numerically and add up to give the solid red lines on the left and right diagram. The solid blue and purple lines on the left and right diagram are respectively the remaining components of Eq.(\ref{Taking the first term of the dielectric susceptibility to dimensionless coordinates for psi < 1}) and Eq.(\ref{The static term of the dielectric susceptibility in dimensionless coordinates for psi > 1}) that have also been solved numerically. Adding the solid-colored lines, we get the solid black lines numbered $(\textbf{I})$ and (\textbf{I$^\prime$}) which are respectively the numerical solutions of Eq.(\ref{Taking the first term of the dielectric susceptibility to dimensionless coordinates for psi < 1}) and Eq.(\ref{The static term of the dielectric susceptibility in dimensionless coordinates for psi > 1}). (b) Comparing the numerical and analytical solutions for the static dielectric response function of a 2D Dirac crystal at T = 0 K. We see that there is good agreement between the two solutions and we also observe a cuspid near the FSN at $\psi\approx1$. (c) Comparing the numerical and analytical solutions for the second addend of the dielectric response function which contains the phonon energy for a 2D Dirac crystal at T = 0 K. We see that there is a good agreement between the two solutions and also observe a strong variation near the FSN at $\psi\approx1$.}
\label{Figure4}
\end{figure}

The integral in Eq.(\ref{The second term of the dielectric response function in dimensionless coordinates}) expresses the second addend of the dielectric response function in Eq.(\ref{Expanding the dielectric susceptibility in terms of w_F}). To derive $\pdv[]{\chi_{0} ({\psi})}{\mathrm{w}_\mathrm{F}}$ of a 2D Dirac crystal we perform the calculations in a polar coordinate system. Therefore, following Eq.(\ref{main equation for dielectric response function}), $\pdv[]{\chi_{0} ({\psi})}{\mathrm{w}_\mathrm{F}}$ of a 2D Dirac crystal at T = 0 K is equal to:
\begin{equation}
\begin{split}
   &-\bigg(\frac{2\pi^2 \mathrm{v}_\mathrm{F}}{e^{2}}\bigg)\pdv[]{\chi_{0} (\psi)}{\mathrm{w}_\mathrm{F}}=\\& \lim_{\epsilon \to 0}\Bigg[\int_{0}^{\frac{\pi}{2}-\epsilon} d\theta \int_{-\frac{\psi}{2}\cos \theta+\frac{1}{2}\sqrt{1-\psi^{2}\sin ^{2}\theta}} ^{\frac{\psi}{2}\cos\theta+\frac{1}{2}\sqrt{1-\psi^{2}\sin ^{2}\theta}} \frac{ \xi \psi d\xi} {2\bigg(\xi^2+\frac{\psi^2}{4}\bigg) \left(1-\sqrt{1-\bigg(\frac{\xi \psi \cos \theta} {\xi^2+\frac{\psi^2}{4}}\bigg)^2}\right)}\Bigg] \\&+\Bigg[\int_{0}^{\arcsin({\frac{1}{\psi_{M}}})} d\theta \int_{\frac{\psi}{2}\cos \theta-\frac{1}{2}\sqrt{1-\psi^{2}\sin ^{2}\theta}} ^{\frac{\psi}{2}\cos\theta+\frac{1}{2}\sqrt{1-\psi^{2}\sin ^{2}\theta}} \frac{ \xi \psi d\xi} {2\bigg(\xi^2+\frac{\psi^2}{4}\bigg) \left(1-\sqrt{1-\bigg(\frac{\xi \psi \cos \theta} {\xi^2+\frac{\psi^2}{4}}\bigg)^2}\right)}\Bigg].
\label{The second term of the dielectric response function in dimensionless coordinates}
\end{split}
\end{equation}
The first integral is for the region $\psi<1$ and the second integral is for the region $\psi>1$. $\psi_{M}$ is the maximum distance we place ourselves from the Fermi sphere to do the calculations for $\psi>1$. Performing the calculations we observe that the second addend of the dielectric response function goes to zero for $\psi>3$ so $\psi_{M}$ can be confidently put equal to 3. To solve the integral in Eq.(\ref{The second term of the dielectric response function in dimensionless coordinates}) analytically we use the binomial expansion: 
\begin{equation}
\begin{split}
   \big(1+\alpha\big)^{n}=1+n\alpha+O(\alpha^2)
\label{binomial expansion}
\end{split}
\end{equation}
Applying the binomial expansion up to the first order of magnitude to the first addend of Eq.(\ref{binomial expansion}) designating the solution for $\psi<1$ we get:
\begin{equation}
\begin{split}
   -\bigg(&\frac{2\pi^2 \mathrm{v}_\mathrm{F}}{e^{2}}\bigg)\pdv[]{\chi_{0} (\psi)}{\mathrm{w}_\mathrm{F}}=\\& \lim_{\epsilon \to 0}\Bigg( \frac{1}{\sqrt{1-\psi^2}}\atanh \Bigg[ \frac{\sqrt{2-2\psi^2} \sin\big(\frac{\pi}{2}-\epsilon\big)} {\sqrt{2-\psi^2+\psi^2 \cos \big(\pi-2\epsilon\big)}}\Bigg]+ \\& \frac{\psi}{2}\tan\big(\frac{\pi}{2}-\epsilon\big) \ln\Bigg[{\frac{2\psi\cos\big(\frac{\pi}{2}-\epsilon\big)+ \sqrt{4-2\psi^2+2\psi^2 \cos \big(\pi-2\epsilon\big)}} {-2\psi\cos\big(\frac{\pi}{2}-\epsilon\big)+ \sqrt{4-2\psi^2+2\psi^2 \cos\big(\pi-2\epsilon\big)}}}\Bigg]\Bigg).
\label{The second term of the dielectric response function for psi<1 in dimensionless coordinates}
\end{split}
\end{equation}
Where Eq.(\ref{The second term of the dielectric response function for psi<1 in dimensionless coordinates}) is the analytical expression of the first integral in Eq.(\ref{The second term of the dielectric response function in dimensionless coordinates}). We have derived the steps to calculate Eq.(\ref{The second term of the dielectric response function for psi<1 in dimensionless coordinates}) in the appendix for completeness. Applying the binomial expansion up to the first order of magnitude to the second addend of Eq.(\ref{The second term of the dielectric response function in dimensionless coordinates}) designating the solution for $\psi>1$ we get:
\begin{equation}
\begin{split}
   -\bigg(&\frac{2\pi^2 \mathrm{v}_\mathrm{F}}{e^{2}}\bigg)\pdv[]{\chi_{0} (\psi)}{\mathrm{w}_\mathrm{F}}=\\&  \frac{1}{\sqrt{1-\psi^2}}\atanh \Bigg[ \frac{\sqrt{2-2\psi^2}} {\psi_{M}\sqrt{2-\psi^2+\psi^2\cos\big(2\arcsin(1/\psi_{M})\big)}}\Bigg]+ \\& \frac{\psi_{M}}{2\sqrt{\psi_{M}^{2}-1}} \psi \ln\Bigg[{\frac{-2\psi\sqrt{1-\frac{1}{\psi_{M}^{2}}}- \sqrt{4-2\psi^2+2\psi^2 \cos\big(2\arcsin(1/\psi_M)\big)}} {2\psi\sqrt{1-\frac{1}{\psi_{M}^{2}}}- \sqrt{4-2\psi^2+2\psi^2 \cos\big(2\arcsin(1/\psi_{M})\big)}}}\Bigg].
\label{The second term of the dielectric response function for psi>1 in dimensionless coordinates}
\end{split}
\end{equation}
Where Eq.(\ref{The second term of the dielectric response function for psi>1 in dimensionless coordinates}) is the analytical expression of the second integral in Eq.(\ref{The second term of the dielectric response function in dimensionless coordinates}). We solve Eq.(\ref{The second term of the dielectric response function in dimensionless coordinates}) numerically and compare it with the analytical results derived in Eq.(\ref{The second term of the dielectric response function for psi<1 in dimensionless coordinates}) and Eq.(\ref{The second term of the dielectric response function for psi>1 in dimensionless coordinates}). The numerical results are shown as the dashed black lines and the analytical results are shown by the red solid line in Fig.~\ref{Figure4}(c). As is shown in Fig.~\ref{Figure4}(c) the two results are in good agreement with one another. By analyzing Fig.~\ref{Figure4}(c) we see that the second-order addend of Eq.(\ref{Expanding the dielectric susceptibility in terms of w_F}) of the dielectric response function diverges at $\psi\approx 1$ making the study of this point important. Higher order terms of the dielectric response function present in  Eq.(\ref{Expanding the dielectric susceptibility in terms of w_F}) can be derived using the same analytical technique. These higher order terms will give us similar plots with divergence at $\psi\approx 1$. However, due to the presence of the term $\mathrm{w_{F}}^{n}$ in the numerator where $\mathrm{w_{F}}<<1$ these higher-order terms can be neglected. By knowing the first and second addend of Eq.(\ref{Expanding the dielectric susceptibility in terms of w_F}) which are respectively the static dielectric response function and the dielectric response function at nonzero phonon energy we can write the total dielectric response function in the two regions of $\psi<1$ and $\psi>1$. For the region $\psi<1$ we insert the values derived for $\chi_0(\psi)$ Eq.(\ref{The static dielectric susceptibility in dimensionless coordinates for psi < 1}), and $\pdv[]{\chi_{0} ({\psi})}{\mathrm{w}_\mathrm{F}}$ Eq.(\ref{The second term of the dielectric response function for psi<1 in dimensionless coordinates}), into Eq.(\ref{Expanding the dielectric susceptibility in terms of w_F}). Therefore, the analytical solution of the total dielectric response function of a 2D Dirac crystal for $\psi<1$ is equal to:
\begin{equation}
\begin{split}
    \chi_{\mathrm{w_{F}}}(\psi)=&
    -\bigg(\frac{e^{2}}{2\pi^2 \mathrm{v}_\mathrm{F}}\bigg)\lim_{\epsilon \to 0} 
    \Bigg(0.59+\frac{3\Big[\pi\psi+\sqrt{1-\psi^{2}} \ln{\Big(\frac{1-\psi-\epsilon}{1+\psi+\epsilon}}\Big)\Big]} {32\psi}+\\& \mathrm{w}_\mathrm{F}\cdot\Bigg(\frac{1}{\sqrt{1-\psi^2}}\atanh \Bigg[ \frac{\sqrt{2-2\psi^2} \sin\big(\frac{\pi}{2}-\epsilon\big)} {\sqrt{2-\psi^2+\psi^2 \cos \big(\pi-2\epsilon\big)}}\Bigg]+ \\& \frac{\psi}{2}\tan\big(\frac{\pi}{2}-\epsilon\big) \ln\Bigg[{\frac{2\psi\cos\big(\frac{\pi}{2}-\epsilon\big)+ \sqrt{4-2\psi^2+2\psi^2 \cos \big(\pi-2\epsilon\big)}} {-2\psi\cos\big(\frac{\pi}{2}-\epsilon\big)+ \sqrt{4-2\psi^2+2\psi^2 \cos\big(\pi-2\epsilon\big)}}}\Bigg]\Bigg)\Bigg).
\label{The total dielectric susceptibility in dimensionless coordinates for psi<1}
\end{split}
\end{equation}
And for the region  $\psi>1$ we insert the values derived for $\chi_0(\psi)$ Eq.(\ref{The static dielectric susceptibility in dimensionless coordinates for psi > 1}), and  $\pdv[]{\chi_{0} ({\psi})}{\mathrm{w}_\mathrm{F}}$ Eq.(\ref{The second term of the dielectric response function for psi>1 in dimensionless coordinates}), into Eq.(\ref{Expanding the dielectric susceptibility in terms of w_F}). Therefore, the analytical solution of the total dielectric response function of a 2D Dirac crystal for $\psi>1$ is:
\begin{equation}
\begin{split}
    \chi_{\mathrm{w_{F}}}(\psi)=&
    -\Bigg(\bigg(\frac{e^{2}}{2\pi^2 \mathrm{v}_\mathrm{F}}\bigg) 
    \Bigg(\frac{0.59}{\psi}+\frac{3\Big[\pi\psi+\sqrt{1-\psi^{2}} \ln{\Big(\frac{1-\psi}{-1+\psi}}\Big)\Big]} {32\psi}\Bigg)+\\& \mathrm{w}_\mathrm{F}\cdot\Bigg(\frac{1}{\sqrt{1-\psi^2}}\atanh\Bigg[ \frac{\sqrt{2-2\psi^2}} {\psi_{M} \sqrt{2-\psi^2+\psi^2 \cos\Big(2\arcsin[{\frac{1}{\psi_{M}}}]\Big)}}\Bigg]+ \\& \frac{\psi_{M}}{2\sqrt{1-\psi_{M}^{2}}}\psi \ln\Bigg[{\frac{-2\psi\sqrt{1-\frac{1}{\psi_{M}^{2}}}+ \sqrt{4-2\psi^2+2\psi^2\cos \Big(2\arcsin[{\frac{1}{\psi_{M}}}]\Big)}} {2\psi\sqrt{1-\frac{1}{\psi_{M}^{2}}}+ \sqrt{4-2\psi^2+2\psi^2\cos \Big(2\arcsin[{\frac{1}{\psi_{M}}}]\Big)}}}\Bigg]\Bigg)\Bigg).
\label{The total dielectric susceptibility in dimensionless coordinates for psi>1}
\end{split}
\end{equation}

We have therefore derived the analytical expression for the dielectric response function of a 2D Dirac crystal in the two regions of $\psi<1$ and $\psi>1$ at T = 0 K. We can derive the dielectric response function of a 2D Dirac crystal for $\mathrm{T}\ne0$ K by employing Eq.(\ref{Dielectric susceptibility at nonzero temperature}) and following the same procedure.

\begin{figure}[h]
\centering
\includegraphics[width=.75\columnwidth]{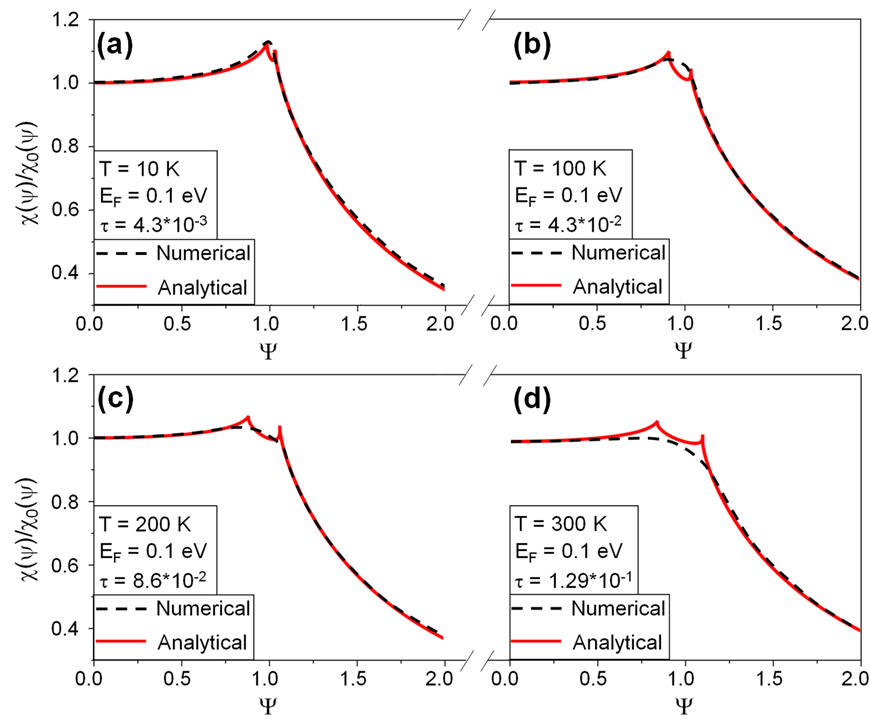}
\caption{The dielectric response function in the 4 graphs (a-d) for different values of $\tau=(k_{B}T/2E_{\mathrm{F}})$ where the solid red lines are the analytical solution, and the dashed black lines are the numerical solution. The analytical solution is done by assuming the Fermi-Dirac distribution function to be a “double step” function while the numerical solution makes no such approximations. We observe that although the analytical solution always has two cuspidal points the numerical solution only has one cuspidal point for $\tau<<1$ and as we increase $\tau$ the single cuspidal point in the numerical solution vanishes. Therefore, the two cuspidal points derived analytically are an artifact of our solution which is caused by using a “double step” function to approximate the Fermi-Dirac distribution.}
\label{Figure5}
\end{figure}

For deriving the dielectric response function at T = 0 K  we approximate the Fermi-Dirac distribution by a “single step” function. However, to derive the dielectric response function at higher temperatures we employ Eq.(\ref{Dielectric susceptibility at nonzero temperature}) where we approximate the Fermi-Dirac distribution with a “double step” function. This results in the dielectric response function having two integrals with their bounds shown by the two purple-shaded regions depicted in Fig.~\ref{Figure2}(c). Each of these two integrals can be solved analytically by following the same procedure we used to solve the integral of the dielectric response function at T = 0 K. To study the dielectric response function for $\mathrm{T}\ne0$ K we also define the variable $\tau = (k_{B}\mathrm{T}/2E_{\mathrm{F}})$ presented in Eq.(\ref{Defiening the dimensionless variables}). The static dielectric response function, $\chi_{0}(\psi)$, of a 2D Dirac crystal derived analytically using the “double step” function for $\mathrm{T}\ne0$ K is shown in Fig.~\ref{Figure5}(a-d) by the solid red lines. We observe that although $\chi_{0}(\psi)$ at T = 0 K has only one cuspidal point, for $\mathrm{T}\ne0$ K the number of cuspidal points increases to two with each integral giving us a separate cuspid. By analyzing Fig.~\ref{Figure5}(a-d) we observe that as we increase the temperature for a given Fermi energy the distance between the cuspidal points increases. This is because as we increase $\tau$ the regions covered by the two integrals in Eq.(\ref{Dielectric susceptibility at nonzero temperature}) get further apart from each other which results in the cuspidal points of the two integrals to also get further apart from one another. To study the nature of the two cuspidal points we derive $\chi_{0}(\psi)$ numerically without applying any approximations to the Fermi-Dirac distribution function and compare it with the analytical solution. The numerical solution is shown in Fig.~\ref{Figure5}(a-d) by the black dashed lines. By analyzing the numerical solution we observe that we only have one cuspidal point for $\tau << 1$ and the cuspidal point starts to vanish as we increase $\tau$. This is not the case for the analytical solution were although as we increase $\tau$ the value of the cuspidal points decrease but they do not vanish. We, therefore, conclude that the two cuspidal points derived analytically are an artifact of our solution which is caused by using a “double step” function to approximate the Fermi-Dirac distribution. By analyzing Fig.~\ref{Figure5}(a-d) we observe that the analytical expression using the “double step” function gives us accurate estimations for $\tau<<1$, and as we increase $\tau$ the analytical solution becomes less accurate, however, it still remains within a good approximation range of the numerical value. This is because although the two cuspids in the analytical solution don’t vanish but their values decrease tending towards the numerical solution.

By deriving the analytical dielectric response function of 2D Dirac crystals as a function of the phonon wave number using the Lindhard model we are able to calculate the phonon line width of these class of condensed matter systems more accurately. To this end, we first write the first order phonon self-energy as a function of $\psi$ as follow:\cite{allen1972neutron}
\begin{equation}
\begin{split}
    \Pi(\psi)= \frac{1}{\sqrt{N_{\psi}}} 
    \sum_{\psi} \big|M_{\psi}\big|^{2}\frac{f_{\boldsymbol{\xi}-\frac{\psi}{2}}-f_{\boldsymbol{\xi}+\frac{\psi}{2}}} {E_{\boldsymbol{\xi}+\frac{\psi}{2}}-{E_{\boldsymbol{\xi}-\frac{\psi}{2}}}-i\,\hbar\omega_{\psi}},
\label{phonon self energy}
\end{split}
\end{equation}
where $\big|M_{\psi}\big|$ is the electron-phonon matrix element\cite{jishi2013feynman}, and $N_\psi$ is the number of points in the summation over $\psi$. The finite line width or the inverse lifetime of a phonon mode is connected to the imaginary part of the phonon self-energy and can be written as:
\begin{equation}
\begin{split}
    \gamma_{\psi}=-2 Im\Pi(\psi)=\frac{2\pi}{\sqrt{N_{\psi}}}\sum_{\psi} \big|M_{\psi}\big|^{2} \Big(f_{\boldsymbol{\xi}-\frac{\psi}{2}}-f_{\boldsymbol{\xi}+\frac{\psi}{2}}\Big)
    \delta\Big(\hbar\omega_{\psi}+{E_{\boldsymbol{\xi}-\frac{\psi}{2}}-{E_{\boldsymbol{\xi}+\frac{\psi}{2}}}}\Big).
\label{phonon line width}
\end{split}
\end{equation}
To calculate the phonon line width we write the electron-phonon matrix element $\big|M_{\psi}\big|$, as a function of the screening potential in the following manner:\cite{jishi2013feynman}
\begin{equation}
\begin{split}
    \big|M_{\psi}\big|=\frac{1}{\Omega}\sum_{\psi}\bigg(\frac{2k_{\mathrm{F}}\hbar^2}{2m\,c\psi}\bigg)^{1/2}\phi_s(\psi)\,\psi,
\label{electron phonon matrix element}
\end{split}
\end{equation}
where $\Omega$ is the unit cell area of the 2D Dirac crystal, m is the ion mass, and $\phi_s(\psi)$ is the screening potential. The screening potential is a function of the dielectric response function and is equal to:\cite{ashcroft2001festkorperphysik}
\begin{equation}
\begin{split}
    \phi_s(\psi)= 
    \Big(\frac{1}{2 k_\mathrm{F}}\Big)\frac{2\pi Q^{2}}{\psi-2\pi\chi(\psi)},
\label{screening potential}
\end{split}
\end{equation}
where $Q$ is the ion charge. By inserting Eq.(\ref{electron phonon matrix element}) and Eq.(\ref{screening potential}) into Eq.(\ref{phonon line width}) we get the phonon line width as a function of $\psi$ in the following manner:
\begin{equation}
\begin{split}
    \gamma_{\psi}=\frac{2\pi^{3}N\hbar^{2}Q^{4}}{mck_{\mathrm{F}}}\frac{1}{\sqrt{N_{\psi}}} \sum_{\psi} \frac{\psi}{\Big(\psi-2\pi\chi(\psi)\Big)^2}\Big(f_{\boldsymbol{\xi}-\frac{\psi}{2}}-f_{\boldsymbol{\xi}+\frac{\psi}{2}}\Big)\delta\Big(c\psi+{E_{\boldsymbol{\xi}-\frac{\psi}{2}}-{E_{\boldsymbol{\xi}+\frac{\psi}{2}}}}\Big).
\label{phonon line width Final}
\end{split}
\end{equation}
We observe that the role of the dielectric response function in the denominator of Eq.(\ref{phonon line width Final}) is essential in determining the phonon line width accurately. By knowing $\chi_{\psi}$, we can write the phonon line width as a function of $\psi$ for acoustic phonons with dispersive energy wavelength relationships. At zero temperature we solve the difference between the Fermi-Dirac distributions using a single-step function and inserting the dielectric response function derived from Eq.(\ref{The total dielectric susceptibility in dimensionless coordinates for psi<1}) and Eq.(\ref{The total dielectric susceptibility in dimensionless coordinates for psi>1}). For nonzero temperatures, we use the double-step function and insert the value of the dielectric response function derived from Eq.(\ref{Dielectric susceptibility at nonzero temperature}) for different temperatures. In the next section, we analyze the dielectric response function and the phonon line width of 2D Dirac crystals extensively. 
\raggedbottom
\section{\label{sec:level4}Discussion}

\begin{figure}[H]
\centering
\includegraphics[width=.75\columnwidth]{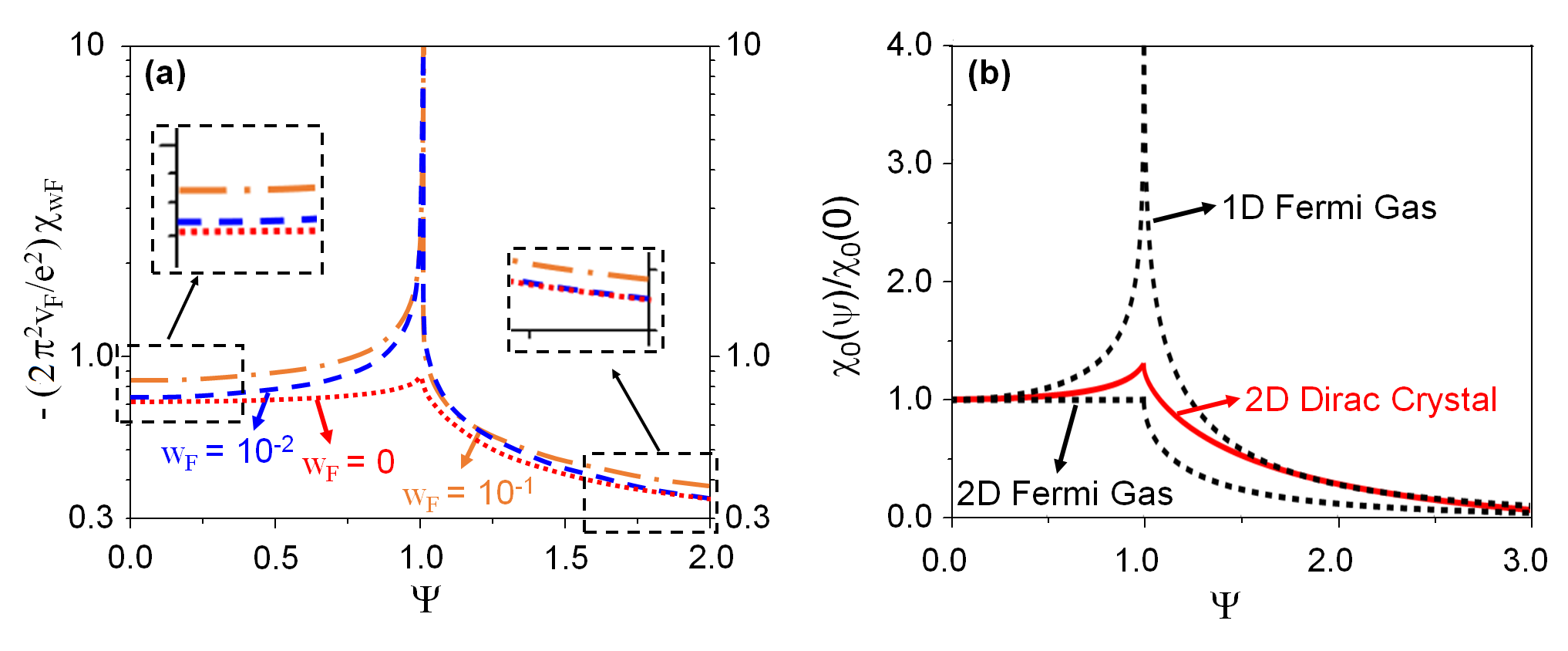}
\caption{(a) The dielectric response function of a 2D Dirac crystal at T = 0 K for different values of $\mathrm{w_{F}}$. We observe that the dielectric response function diverges near the FSN at the point $\psi\approx1$ even for small values of $\mathrm{w_{F}}\ne0$. The negligible difference in the dielectric response function of various 2D Dirac crystals with different values of $\mathrm{w_{F}}$  shows the generality of our solution for 2D Dirac crystals. (b) Comparing $\chi_{0}(\psi)$ of a 2D Dirac crystal with that of a 1D and 2D Fermion gas at T = 0 K. We observe that $\chi_{0}(\psi)$ of a 2D Dirac crystal does not follow a constant line like the 2D Fermi gas and it increases reaching a maximum at the FSN point, making the study of 2D Dirac crystals more intricate.}
\label{Figure6}
\end{figure}

We have analytically derived the dynamic dielectric response function of electrons strongly correlated to acoustic phonons in 2D Dirac crystals as a function of the phonon wave number, Eqs.(\ref{The total dielectric susceptibility in dimensionless coordinates for psi<1},\ref{The total dielectric susceptibility in dimensionless coordinates for psi>1}). Our expression of the dielectric response function is general enough to describe the electron-lattice interaction at any Fermi-level shifts and temperatures even for cases where the Sommerfeld approximation is not valid. Also, as can be seen from Eqs.(\ref{The total dielectric susceptibility in dimensionless coordinates for psi<1},\ref{The total dielectric susceptibility in dimensionless coordinates for psi>1}), our expression of the dielectric response function has the advantage of being applicable to various 2D Dirac crystals with different Fermi velocities, $\mathrm{v_{F}}$, and not only to a specific type of 2D Dirac crystal. The analytical expression of the dielectric response function derived in our paper for acoustic phonons where $\hbar \omega_{q}=cq$ is more accurate compared to methods that use the RPA method. For example, using the RPA Wusch et al.\cite{wunsch2006dynamical} states that at $q = 2k_\mathrm{F}$ or $\psi=1$ the dielectric response function has a discontinuity only in the second derivative while we show that in the static regime, the dielectric response function has a discontinuity in the first derivative and in the dynamic regime the dielectric response function diverges itself at $\psi=1$, Fig.~\ref{Figure6}(a). A literature review of the work done on the dielectric response function in 2D Dirac crystals is shown in Table 1. Furthermore comparing our results to the Lindhard-Mermin dielectric function\cite{mermin1970lindhard} we observe that using RPA Mermin derives the Lindhard dielectric function in the relaxation time approximation by assuming the relaxation to occur not towards the thermal equilibrium but to a more general distribution characterized by the chemical potential $\mu$. The advantage of the Mermin approach is its simplicity, however, it demands further systematic elaboration of the quantum statistical theory of the dielectric response function and a more comprehensive formulation of the chemical potential beyond the Sommerfeld approximation for 2D Dirac crystals. Also, our analytical expression of the dielectric response function written as a function of the phonon wave number is necessary to better understand some of the features of 2D Dirac crystals such as the phonon self-energy. The customarily used model to derive the phonon self-energy is the jellium model\cite{jishi2013feynman}. Jellium models disregard the lattice structure and only consider free electrons. Furthermore, the ions in the jellium are approximated into a uniform background of positive charges resulting in a constant electrostatic potential. Such a model is suitable for studying electron-electron and electron-phonon interactions in metals but not in 2D Dirac crystals where the electrostatic potential undergoes strong fluctuations. Specifically, in 2D Dirac crystals the electronic dispersion is linear (i.e. $E=\mathrm{v}_{\mathrm{F}}k$ as opposed to $E= \hbar^{2}k^{2}/2m$ as in the jellium). In order to consider this, the electrostatic potential is assumed to be Lindhard-screened, as in Eq.(\ref{screening potential}). This approach, which also considers the specific crystalline structure of 2D Dirac crystals leading to Fermi-level electrons at the Brillouin-zone K-point, has been used by us to derive the phonon line width in our model as in Eq.(\ref{phonon line width Final}). It is worthwhile noting that the non-uniformity of the electrostatic potential plays a critical role in determining the phonon line width at wave numbers $q$ approaching Fermi surface nesting conditions, where we anticipate it to increase linearly with the temperature in agreement with the experimental results \cite{debernardi2002anharmonic}. This would not be the case if the electron-phonon coupling would have been considered at the approximation level of the jellium model.
\begin{table}[h]
\begin{tabular}{lll}
\hline
\multicolumn{1}{|c|}{Literature}                                                                & \multicolumn{1}{c|}{Reported Results}                                                                                                                                                                                                                                & \multicolumn{1}{c|}{Limitations}                                                                                                                                                                                                           \\ \hline
\multicolumn{1}{|l|}{\begin{tabular}[c]{@{}l@{}}Hwang {[}31{]}\\ Bahrami {[}32{]}\end{tabular}} & \multicolumn{1}{l|}{\begin{tabular}[c]{@{}l@{}}Have calculated the dielectric response \\ function by assuming phonons to have \\ non-dispersive energy-wavelengths\end{tabular}}                                                                                    & \multicolumn{1}{l|}{\begin{tabular}[c]{@{}l@{}}Have neglected phonons with dispersive energy \\ wavelengths.\end{tabular}}                                                                                                                  \\ \hline
\multicolumn{1}{|l|}{Iurov {[}34{]}}                                                            & \multicolumn{1}{l|}{\begin{tabular}[c]{@{}l@{}}Have calculated the dielectric response\\ function for silicene illuminated by \\ circularly polarized light.\end{tabular}}                                                                                           & \multicolumn{1}{l|}{\begin{tabular}[c]{@{}l@{}}The disturbances studied are, photons with fixed \\ energy and not phonons and the study on silicene \\ cannot be generalized to other Dirac crystals.\end{tabular}}                         \\ \hline
\multicolumn{1}{|l|}{Lu {[}35{]}}                                                               & \multicolumn{1}{l|}{\begin{tabular}[c]{@{}l@{}}Have calculated the zero-temperature,\\ static, Lindhard response function for \\ single-layer and bilayer graphene.\end{tabular}}                                                                                    & \multicolumn{1}{l|}{\begin{tabular}[c]{@{}l@{}}Only assumes the static dielectric response function \\ at zero temperatures. Also, the study on graphene \\ cannot be generalized to other Dirac crystals.\end{tabular}}                    \\ \hline
\multicolumn{1}{|l|}{Calandra {[}36{]}}                                                         & \multicolumn{1}{l|}{\begin{tabular}[c]{@{}l@{}}Have calculated the e-ph coupling in \\ electron-doped graphene using e-ph \\ matrix elements extracted from density \\ functional theory simulations.\end{tabular}}                                                  & \multicolumn{1}{l|}{\begin{tabular}[c]{@{}l@{}}A direct analytical calculation has not been made \\ and the results can not be extended beyond the \\ specific crystal under study.\end{tabular}}                                           \\ \hline
\multicolumn{1}{|l|}{Wunsch {[}30{]}}                                                           & \multicolumn{1}{l|}{\begin{tabular}[c]{@{}l@{}}Have used RPA to calculate the response \\ the function of graphene in the two \\ scenarios of $q\to0$ relevant for photon \\ spectroscopy and $\hbar \omega_{q}=0$  \\ applicable for the static case.\end{tabular}} & \multicolumn{1}{l|}{\begin{tabular}[c]{@{}l@{}}A direct analytical calculation has not been made, \\ phonon wave functions where $q \ne 0$ have not \\ been considered and the results cannot be \\ extended beyond graphene.\end{tabular}} \\ \hline
\multicolumn{1}{|l|}{Lazzeri {[}27{]}}                                                          & \multicolumn{1}{l|}{\begin{tabular}[c]{@{}l@{}}Have derived the dielectric response \\ function in doped graphene as a function \\ of the charge doping for $q = 0$\end{tabular}}                                                                                    & \multicolumn{1}{l|}{\begin{tabular}[c]{@{}l@{}}Phonon wave functions where $q \ne 0$ have not \\ been considered, and the results cannot be \\ extended beyond graphene.\end{tabular}}                                                      \\ \hline
                                                                                                &                                                                                                                                                                                                                                                                      &                                                                                                                                                                                                                                            
\end{tabular}
\caption{\label{demo-table}Literature review on the dielectric response function of 2D Dirac Crystals.}
\end{table}

We will now analyze the dynamic dielectric response function along the $\psi$ axis and discuss the physical meaning of the observed cuspid and divergence near the FSN region. As shown in Eq.(\ref{Expanding the dielectric susceptibility in terms of w_F}) the dielectric response function of a 2D Dirac crystal is the sum of the static term with zero phonon energy, $\mathrm{w_{F}}=0$, and higher order terms where the phonon energy is nonzero, $\mathrm{w_{F}}\ne0$. In Fig.~\ref{Figure6}(a) we plot the dielectric response function of a 2D Dirac crystal as a function of $\psi$ at $\mathrm{T}=0$ K for different values of $\mathrm{w_{F}}$. By analyzing Fig.~\ref{Figure6}(a) and comparing the red dashed line where $\mathrm{w_{F}}=0$, with the blue and orange dashed lines where $\mathrm{w_{F}}\ne0$, we observe that at the two regions $\psi<1$ and $\psi>1$, the static term is dominant, however, near the FSN region, $\psi\approx1$, the higher order terms become more dominant. We observe that the dielectric response function diverges near the FSN region at the point $\psi\approx1$ even for small values of $\mathrm{w_{F}}\ne0$ where the speed of sound is small, albeit non-negligible, over the Dirac-electron Fermi velocity. It is only when $\mathrm{w_{F}}=0$ and we assume that we are in a static regime that the strong variation in the dielectric response function vanishes, and we only have a cuspid. The divergence of the dielectric response function in the dynamic regime and the cuspid in the static regime near the FSN region at $\tau\approx0$ is a consequence of the particular enhanced electron-phonon interaction at $q\approx 2k_{\mathrm F}$ or $\psi \approx 1$. However, in the static regime, we observe that the cuspid vanishes as we increase $\tau$ as shown in Fig.~\ref{Figure5}. This is because, as we increase the temperature more electrons begin to get thermally excited to energy levels close to $E_\mathrm{F}$ smearing out the "single step" Fermi-Dirac approximation making the particular enhanced electron-phonon interaction more general which smooths out the cuspid observed in Fig.~\ref{Figure5}(a). It should be noted that due to the strong variations in the dynamic dielectric response function near the FSN region of 2D Dirac crystals as shown in Fig.~\ref{Figure6}(a), it is important not to simplify the problem to the static case where the energy of acoustic phonons is put equal to zero. Such physical settings can be the study of the thermal and electrical variations of 2D Dirac crystals caused by electron-phonon interactions\cite{kazemian2017modelling}. Also, when studying the higher order terms of electron-phonon interactions in 2D Dirac crystals with the employment of a Fermion propagator,\cite{jishi2013feynman,mandl2010quantum} which has a time component, we have to consider the energy of the phonons and cannot work in the static regime.\cite{kazemian2017modelling} By further analyzing Fig.~\ref{Figure6}(a) we observe that the difference in the dielectric response function of various 2D Dirac crystals with different values of $\mathrm{w_{F}}$\cite{xu2016hydrogenated,hwang2012fermi,lee2015fermi,kara2012review,diaz2017tuning}, is negligible showing the generality of our solution.

We will further discuss the effect of the disorder on the dielectric response function near the FSN region for different values of $k_{\mathrm{F}}$. Disorder affects 2D Dirac systems differently at different values of $k_{\mathrm{F}}$ and $q$ introducing $\delta k_{\mathrm{F}}$ and $\delta q$ both of the order of $1/L_{dis}$ (with $L_{dis}$ being the order range of the disorder). Studying our expression of the dielectric response function written as a function of $\psi$, disorder significantly affects our considerations when $\Delta \psi<\delta \psi$. $\Delta \psi$ is the full-width height-maximum (FWHM) of the dielectric response function near the FSN region and as can be seen from Fig.~\ref{Figure6}(a) in the static case (red dashed line) $\Delta \psi_{stat} \approx 0.2$ and in the dynamic case (blue dashed line) $\Delta \psi_{dyn} \approx 0.1$. We derive $\delta \psi$ by taking the partial derivative of $\psi$ as follow:
\begin{equation}
    \delta \psi=\Big|\pdv[]{\psi}{q} \Big|\delta q+\Big|\pdv[]{\psi}{k_\mathrm{F}}\Big|\delta k_\mathrm{F}=\frac{1}{k_\mathrm{F}}\Big(\frac{\delta q}{2}+\psi \cdot \delta k_{\mathrm{F}}\Big).
\label{delta psi as a function of q and k_F}
\end{equation}
We assume $\delta k_{\mathrm{F}}=\delta q \approx 1/L_{dis}$ and $\psi \approx 1$ near the FSN region. We therefore have:
\begin{equation}
    \delta \psi \approx \frac{3}{2k_{\mathrm{F}}} \frac{1}{L_{dis}}.
\label{delta psi as a function of k_F and L_dis}
\end{equation}
To study the different values of $k_\mathrm{F}$ where the disorder becomes important we assume the disorder range to be $L_{dis}=10^{-5}m$. For $k_{\mathrm{F}}=10^{8}m^{-1}$ we have $\delta \psi = 0.001$. Therefore, $\delta \psi$ is smaller than both $\Delta \psi_{stat}$ and $\Delta \psi_{dyn}$ resulting in the effect of the disorder to be negligible. Making the value of $k_{\mathrm{F}}$ smaller for $k_{\mathrm{F}}=10^{6}m^{-1}$ we have $\delta \psi = 0.15$. In this scenario for the static dielectric response function, we have $\delta \psi<\Delta \psi_{stat}=0.2$, and although the disorder is significant our results are still reliable. However, for the dynamic dielectric response function, we have $0.1=\Delta \psi_{dyn}<\delta \psi$ and our results become unreliable. From Eqs.(\ref{delta psi as a function of k_F and L_dis} we observe that as $k_\mathrm{F} \to 0$, the role of the disorder becomes more significant and in order to counterbalance it, the length of the disorder range should approach infinity, $L_{dis} \to \infty$.

We further compare the static dielectric response function, $\chi_{0}(\psi)$, of a 2D Dirac crystal with that of a Fermion gas at T = 0 K \cite{mihaila2011lindhard}. This is shown in Fig.~\ref{Figure6}(b). We observe that $\chi_{0}(\psi)$ of a 2D Dirac crystal is reminiscent of that of a 1D and a 2D Fermion gas. As can be seen from Fig.~\ref{Figure6}(b) $\chi_{0}(\psi)$ of a 2D Dirac crystal lies between that of a 1D and 2D Fermion gas not diverging near the FSN region like the former and not approaching it along a constant line like the latter. Hence, when studying $\chi_{0}(\psi)$ of a 2D Dirac crystal it is important to distinguish it from a 2D Fermi gas. While $\chi_{0}(\psi)$ of a 2D Fermi gas approaches the FSN along a constant line, for 2D Dirac crystals $\chi_{0}(\psi)$ increases reaching a maximum at $\psi=1$. This makes the study of $\chi_{0}(\psi)$ for 2D Dirac crystals more intricate in the FSN region. By further analyzing Fig.~\ref{Figure6}(b) we observe that in the region of $\psi>>1$, $\chi_{0}(\psi)$ of the Fermi gas and 2D Dirac crystals are both proportional to $1/\psi$ which is equivalent to the result we get when we use the Thomas-Fermi approach. Therefore, the Thomas-Fermi model for calculating the dielectric response function at long wavelengths is equal for 2D Dirac crystals and Fermi gasses.

We are able to solve the phonon line width, Eq.(\ref{phonon line width Final}) numerically for different 2D Dirac crystals by performing the summation over $\psi$. To analyze the correlation between the phonon line width and the phonon wave number we derive $\gamma_{\psi}$ for different values of $\psi$ within the range of $0.2<\psi<1$. Fig. 7(a) shows that there is a linear correlation between $\gamma_{\psi}$ and $\psi$, with the linear fit having the coefficient of determination $R^2=0.99$. We, therefore, conclude that similar to the linear proportionality between the spectral energy and the fermion wave number, $E\propto k$, there is also a linear correlation between the phonon line width and the phonon wave number, $\gamma_{q}\propto q$, in 2D Dirac crystals. We further compare the phonon line width approaching the FSN condition of 2D Dirac crystals at different temperatures in the static scenario. By studying Fig. 7(b) we observe a linear increase in the phonon line width or a linear decrease in the lifetime of the phonon mode as we increase the temperature. This suggests that at higher temperatures where the lattice vibrations increase the lifetime of the phonon mode decreases. Our results further match the experimental data collected on germanium \cite{debernardi2002anharmonic} which further confirms our results.
\begin{figure}[H]
\centering
\includegraphics[width=.75\columnwidth]{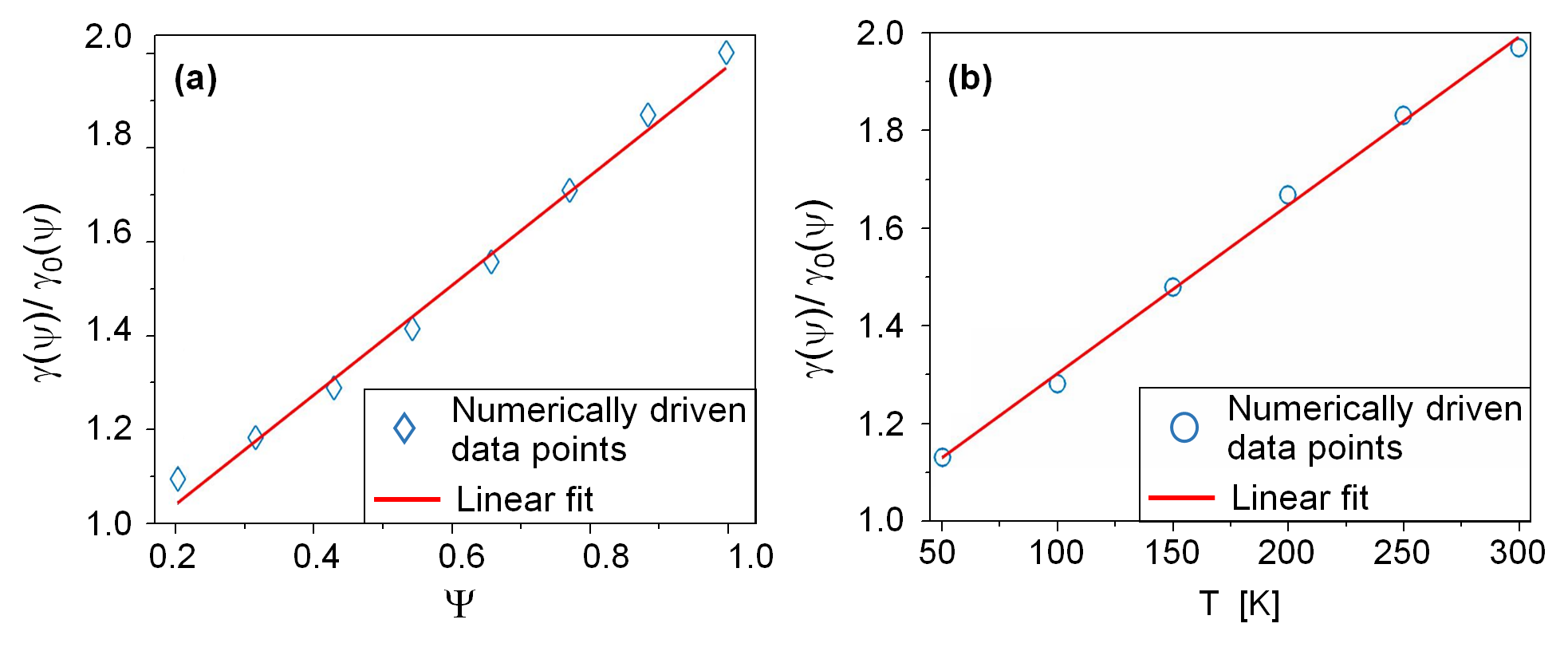}
\caption{(a) the phonon line width vs $\psi$ in 2D Dirac crystals. The linear fit has the coefficient of determination $R^2=0.99$, indicating a linear correlation between the phonon line width and the phonon wave number. (b) The phonon line width near the FSN of 2D Dirac crystals at different temperatures. The linear fit has the coefficient of determination $R^2=0.99$, indicating a linear correlation between the phonon line width and the lattice temperature.}
\label{Figure7}
\end{figure}

\section{\label{sec:level5}Conclusion}

In this paper by using the Lindhard model we derive an analytical expression for the dielectric response function of electrons strongly correlated to acoustic phonons in 2D crystals. The dielectric response function derived is a function of the phonon wave number which helps us better understand phenomenons in 2D Dirac crystals such as electron-phonon interactions and phonon self-energy. We see that different from free-electron systems, in 2D Dirac crystals the dielectric response function exhibits a cuspidal point near the FSN region in the static case where the phonon energy has been put equal to zero. We further observe when the phonon energy has not been put equal to zero the dielectric response function of a 2D Dirac crystal varies strongly near the FSN region even when the speed of sound is small, albeit nonnegligible, over the Dirac-electron Fermi velocity. We, therefore, show that the common approach for calculating the dielectric response function of 2D Dirac crystals by putting the energy of the phonons equal to zero is not accurate. We also show the generality of our solution for different 2D Dirac crystals with different Fermi velocities. We further use the dielectric response function obtained for 2D Dirac crystals to derive their phonon line width. We observe that the phonon line width increases as we move towards the FSN and as we go towards higher temperatures which matches the experimental work presented in the literature. 

\centerline{***}
\paragraph*{\bf Acknowledgments}
We would like to thank Dr. Ghazal Farhani for helping us with the numerical calculations of equations (11) and (16) and the discussions on the phonon self energy. Mr. Victor Wong is acknowledged for pointing out several references of this paper. This work was supported by the Natural Science and Engineering Research Council of Canada (NSERC) through a Discovery Grant (RGPIN-2020-06669). We acknowledge the Anishinaabek, Haudenosaunee, L¯unaap´eewak and Attawandaron peoples, on whose traditional lands Western University is located.%

\appendix

\section{Appendix: Detailed calculation of the second addend of the dielectric response function}

The second addend of the dielectric response function at zero temperature for $\psi<1$ is:
\setcounter{equation}{0}
\renewcommand\theequation{A.\arabic{equation}}
\begin{equation}
\begin{split}
   &-\bigg(\frac{\pi^2 \mathrm{v}_\mathrm{F}}{e^{2}}\bigg)\frac{1}{2k_\mathrm{F}}\pdv[]{\chi_{0} (\psi)}{\mathrm{w}_\mathrm{F}}=\\& \lim_{\epsilon \to 0}\Bigg[\int_{0}^{\frac{\pi}{2}-\epsilon} d\theta \int_{-\frac{\psi}{2}\cos \theta+\frac{1}{2}\sqrt{1-\psi^{2}\sin ^{2}\theta}} ^{\frac{\psi}{2}\cos\theta+\frac{1}{2}\sqrt{1-\psi^{2}\sin ^{2}\theta}} \frac{ \xi \psi d\xi} {2\bigg(\xi^2+\frac{\psi^2}{4}\bigg) \left(1-\sqrt{1-\bigg(\frac{\xi \psi \cos \theta} {\xi^2+\frac{\psi^2}{4}}\bigg)^2}\right)}\Bigg].
\label{Detailed calculations of the second addend term of the dielectric response function for psi<1}
\end{split}
\end{equation}
To solve the integral analytically we use the binomial expansion. Using the binomial expansion up to the first order the denominator can be written as:
\begin{equation}
\begin{split}
   2(\xi^{2}+\frac{\psi^{2}}{4})(1-1+\frac{1}{2}\frac{\psi^2 \xi^2 \cos{\theta}^{2}}{(\xi^2+\frac{\psi^2}{4})^2})=\frac{\psi^2 \xi^2 \cos{\theta}^{2}}{(\xi^2 + \frac{\psi^2}{4})}.
\label{Deriving the denominator of the second addend of the dielectric response function for psi<1 using the binomial expansion}
\end{split}
\end{equation}
We therefore have:
\begin{equation}
\begin{split}
   &\lim_{\epsilon \to 0}\Bigg(\int_{0}^{\frac{\pi}{2}-\epsilon} d\theta \int_{-\frac{\psi}{2}\cos \theta+\frac{1}{2}\sqrt{1-\psi^{2}\sin ^{2}\theta}} ^{\frac{\psi}{2}\cos\theta+\frac{1}{2}\sqrt{1-\psi^{2}\sin ^{2}\theta}} d\xi \frac{(\xi^2 + \frac{\psi^2}{4})}{\psi \xi \cos{\theta}^{2}} \Bigg)=\\&\lim_{\epsilon \to 0}\Bigg(\int_{0}^{\frac{\pi}{2}-\epsilon} \frac{d\theta}{2\psi \cos{\theta}^{2}}\Bigg[\psi^{2}\ln\Bigg({\frac{\psi\cos\theta+\sqrt{1-\psi^{2}\sin ^{2}\theta}} {-\psi\cos\theta+\sqrt{1-\psi^{2}\sin ^{2}\theta}}}\Bigg)+\Bigg(2\psi\cos{\theta\sqrt{1-\psi^{2}\sin{\theta}^{2}}}\Bigg)\Bigg]\Bigg)
\label{Deriving second addend of the dielectric response function for psi<1 over xi}
\end{split}
\end{equation}
Finally integrating Eq.(\ref{Deriving second addend of the dielectric response function for psi<1 over xi}) over $\theta$ we have:
\begin{equation}
\begin{split}
   \lim_{\epsilon \to 0}&\Bigg(\int_{0}^{\frac{\pi}{2}-\epsilon} \frac{d\theta}{2\psi \cos{\theta}^{2}}\Bigg[\psi^{2}\ln\Bigg({\frac{\psi\cos\theta+\sqrt{1-\psi^{2}\sin ^{2}\theta}} {-\psi\cos\theta+\sqrt{1-\psi^{2}\sin ^{2}\theta}}}\Bigg)+\Bigg(2\psi\cos{\theta\sqrt{1-\psi^{2}\sin{\theta}^{2}}}\Bigg)\Bigg]\Bigg)\\& =\lim_{\epsilon \to 0}\Bigg( \frac{1}{\sqrt{1-\psi^2}}\atanh \Bigg[ \frac{\sqrt{2-2\psi^2} \sin\big(\frac{\pi}{2}-\epsilon\big)} {\sqrt{2-\psi^2+\psi^2 \cos \big(\pi-2\epsilon\big)}}\Bigg]+\\& \frac{\psi}{2}\tan\big(\frac{\pi}{2}-\epsilon\big) \ln\Bigg[{\frac{2\psi\cos\big(\frac{\pi}{2}-\epsilon\big)+ \sqrt{4-2\psi^2+2\psi^2 \cos \big(\pi-2\epsilon\big)}} {-2\psi\cos\big(\frac{\pi}{2}-\epsilon\big)+ \sqrt{4-2\psi^2+2\psi^2 \cos\big(\pi-2\epsilon\big)}}}\Bigg]\Bigg).
\label{The second term of the dielectric response function for psi<1 in dimensionless coordinates}
\end{split}
\end{equation}
The same derivations can be made for $\psi>1$.

\bibliographystyle{unsrt}
{\footnotesize
\bibliography{PINN.bib}}

\end{document}